\documentclass[12pt,english,journal,draftcls,onecolumn,12pt,twoside]{IEEEtran}
\usepackage[T1]{fontenc}
\usepackage[latin9]{inputenc}
\usepackage{babel}
\usepackage{verbatim}
\usepackage{varioref}
\usepackage{amsmath}
\usepackage{amsthm}
\usepackage{amssymb}
\usepackage{graphicx}
\usepackage{esint}
\usepackage[unicode=true,pdfusetitle,
 bookmarks=true,bookmarksnumbered=false,bookmarksopen=false,
 breaklinks=false,pdfborder={0 0 1},backref=false,colorlinks=false]
 {hyperref}

\makeatletter

\providecommand{\tabularnewline}{\\}

\theoremstyle{plain}
\newtheorem{thm}{\protect\theoremname}
\theoremstyle{definition}
\newtheorem{defn}[thm]{\protect\definitionname}
\theoremstyle{plain}
\newtheorem{lem}[thm]{\protect\lemmaname}
\theoremstyle{plain}
\newtheorem{prop}[thm]{\protect\propositionname}
\theoremstyle{plain}
\newtheorem{cor}[thm]{\protect\corollaryname}
\theoremstyle{remark}
\newtheorem{rem}[thm]{\protect\remarkname}
\theoremstyle{remark}
\newtheorem{claim}[thm]{\protect\claimname}

\usepackage{babel}
\usepackage{notoccite}

\providecommand{\claimname}{Claim}
\providecommand{\corollaryname}{Corollary}
\providecommand{\definitionname}{Definition}

\providecommand{\lemmaname}{Lemma}
\providecommand{\propositionname}{Proposition}
\providecommand{\theoremname}{Theorem}

\makeatother

\providecommand{\claimname}{Claim}
\providecommand{\corollaryname}{Corollary}
\providecommand{\definitionname}{Definition}
\providecommand{\lemmaname}{Lemma}
\providecommand{\propositionname}{Proposition}
\providecommand{\remarkname}{Remark}
\providecommand{\theoremname}{Theorem}

\begin{document}
\global\long\def\expect#1{\mathbb{E}\left[#1\right]}

\global\long\def\abs#1{\left\lvert #1\right\lvert }

\global\long\def\brac#1{\left(#1\right)}

\global\long\def\lgbrac#1{\log\left(#1\right)}

\global\long\def\lnbrac#1{\ln\left(#1\right)}

\global\long\def\lghbrac#1{\log\left(2\pi e\brac{#1}\right)}

\global\long\def\cbrac#1{\left\{  #1\right\}  }

\global\long\def\sbrac#1{\left[#1\right] }

\global\long\def\Det#1{\left|#1\right|}

\global\long\def\rline#1{\left.#1\right| }

\global\long\def\prob#1{\mathbb{P}\brac{#1}}

\global\long\def\union{\bigcup}

\global\long\def\inter{\bigcap}

\global\long\def\real{\mathbb{R}}

\global\long\def\pderiv#1#2{\frac{\partial#1}{\partial#2}}

\global\long\def\etal{{\it et al.}~}

\newcounter{mytempeqncnt}

\allowdisplaybreaks

\newif\ifarxiv

\arxivtrue

\title{Approximate Capacity of Fast Fading Interference Channels with No
Instantaneous CSIT \thanks{Shorter versions of this work appeared in \cite{Joyson_fading,Joyson_AF_isit}
with outline of proofs. This version has complete proofs. This work
was supported in part by NSF grants 1514531, 1314937 and a gift from
Guru Krupa Foundation.}}

\author{Joyson Sebastian, Can Karakus, Suhas Diggavi}
\maketitle
\begin{abstract}
We develop a characterization of fading models, which assigns a number
called \textit{logarithmic Jensen's gap} to a given fading model.
We show that as a consequence of a finite logarithmic Jensen's gap,
approximate capacity region can be obtained for fast fading interference
channels (FF-IC) for several scenarios. We illustrate three instances
where a constant capacity gap can be obtained as a function of the
logarithmic Jensen's gap. Firstly for an FF-IC with neither feedback
nor instantaneous channel state information at transmitter (CSIT),
if the fading distribution has finite logarithmic Jensen's gap, we
show that a rate-splitting scheme based on average interference-to-noise
ratio  ($inr$) can achieve its approximate capacity. Secondly we
show that a similar scheme can achieve the approximate capacity of
FF-IC with feedback and delayed CSIT, if the fading distribution has
finite logarithmic Jensen's gap. Thirdly, when this condition holds,
we show that point-to-point codes can achieve approximate capacity
for a class of FF-IC with feedback. We prove that the logarithmic
Jensen's gap is finite for common fading models, including Rayleigh
and Nakagami fading, thereby obtaining the approximate capacity region
of FF-IC with these fading models. %
\end{abstract}

\section{Introduction}

The 2-user Gaussian IC is a simple model that captures the effect
of interference in wireless networks. Significant progress has been
made in the last decade in understanding the capacity of the Gaussian
IC \cite{han_kobayashi,chong2008han,etkin_tse_no_fb_IC,suh_tse_fb_gaussian}.
In practice the links in the channel could be time-varying rather
than static. Characterizing the capacity of FF-IC without CSIT has
been an open problem. In this paper we make progress in this direction
by obtaining the capacity region of certain classes of FF-IC without
instantaneous CSIT within a constant gap.

\subsection{Related work}

Previous works have characterized the capacity region to within a
constant gap for the IC where the channel is known at the transmitter
and receiver. The capacity region of the 2-user IC without feedback
was characterized to within 1 bit/s/Hz in \cite{etkin_tse_no_fb_IC}.
In \cite{suh_tse_fb_gaussian}, Suh and Tse characterized the capacity
region of the IC with feedback to within 2 bits per channel use. These
results were based on the Han-Kobayashi scheme \cite{han_kobayashi},
where the transmitters use superposition coding splitting their messages
into common and private parts, and the receivers use joint decoding.
Other variants of wireless networks based on the IC model have been
studied in literature. The interference relay channel (IRC), which
is obtained by adding a relay to the 2-user interference channel (IC)
setup, was introduced in \cite{sahin_Erkip_IRC_2007} and was further
studied in \cite{Tiann_Yener_IRC_2011,Maric_Dabora_Goldsmith_IRC_2012,Bassi_Piantanida_Yang_IRC_2015,Gherekhloo_Chaaban_Sezgin_IRC_2016}.
In \cite{Wang_Tse_IC_Rx_coop}, Wang and Tse studied the IC with receiver
cooperation. The IC with source cooperation was studied in \cite{Prabhakaran_IC_source_cooperation,Wang_Tse_IC_Tx_coop}.

{} When the channels are time varying, most of the existing techniques
for IC cannot be used without CSIT.%
{} In \cite{Farsani_fading}, Farsani showed that if each transmitter
of FF-IC has knowledge of the $inr$ to the non-corresponding receiver\footnote{For Tx$1$ the non-corresponding receiver is Rx$2$ and similarly
for Tx$2$ the non-corresponding receiver is Rx$1$ }, the capacity region can be achieved within 1 bit/s/Hz. Lalitha \etal
\cite{Lalitha_Ergodic_fading} derived sum-capacity results for a
subclass of FF-IC with perfect CSIT. The idea of interference alignment
\cite{cadambe_jafar_interference_alignment} has been extended to
FF-IC to obtain the degrees of freedom (DoF) region for certain cases.
The degrees of freedom region for the MIMO interference channel with
delayed CSIT was studied in \cite{MIMO_IC_delayed_csit_Vaze_2012}.
Their results show that when \emph{all users have single antenna},
the DoF region is same for the cases of no CSIT, delayed CSIT and
instantaneous CSIT. The results from \cite{MIMO_IC_FB_delayed_CSIT_Tandon_2013}
show that DoF region for FF-IC with output feedback and delayed CSIT
is contained in the DoF region for the case with instantaneous CSIT
and no feedback. Kang and Choi \cite{ergodic_delayed_kang2013} considered
interference alignment for the K-user FF-IC with delayed channel state
feedback and showed a result of $\frac{2K}{K+2}$ DoF. They also showed
the same DoF can be achieved using a scaled output feedback, but without
channel state feedback. Therefore, the above works have characterizations
for DoF for several fading scenarios, and also show that for single
antenna systems, feedback is not very effective in terms of DoF. However,
as we show in this paper, the situation changes when we look for more
than DoF, and for approximate optimality of the entire capacity region.
In particular, we allow for arbitrary channel gains, and do not limit
ourselves to SNR-scaling results\footnote{However, we can also use our results to get the \emph{generalized}
DoF studied in \cite{etkin_tse_no_fb_IC} for the FF-IC. This shows
that for generalized DoF, feedback indeed helps, as shown in our results.}. In particular, we show that though the capacity region is same (within
a constant) for the cases of no CSIT, delayed CSIT and instantaneous
CSIT, there is a significant difference with output feedback. When
there is output feedback and delayed CSIT the capacity region is larger
than that for the case with no feedback and instantaneous CSIT in
contrast to the DoF result from \cite{MIMO_IC_FB_delayed_CSIT_Tandon_2013}.
This gives us a finer understanding of the role of CSIT as well as
feedback in FF-IC with arbitrary (and potentially asymmetric) link
strengths, and is one of the main contributions of this paper.

Some simplified fading models have been introduced to characterize
the capacity region of the FF-IC in the absence of CSIT. In \cite{wang2013bursty},
Wang \etal considered the bursty IC, where the presence of interference
is governed by a Bernoulli random state. The capacity of one-sided
IC under ergodic layered erasure model, which considers the channel
as a time-varying version of the binary expansion deterministic model
\cite{avest_det}, was studied in \cite{aggarwal_fading_Z_2009,zhu_fading_z_2011}.
The binary fading IC, where the channel gains, the transmit signals
and the received signals are in the binary field was studied in \cite{vahid2014binaryfading,vahid2017binaryfading_no_CSIT}
by Vahid \etal In spite of these efforts, the capacity region of
FF-IC without CSIT is still unknown, and this paper presents what
we believe to be the first approximate characterization of the capacity
region of FF-IC without CSIT, for a class of fading models satisfying
the regularity condition, defined as the finite \emph{logarithmic
Jensen's gap}.

{}

\subsection{Contribution and outline}

In this paper we first introduce the notion of \emph{logarithmic Jensen's
gap} for fading models. This is defined in Section \ref{sec:fading_models}
as a number calculated for a fading model depending on the probability
distribution for the channel strengths. It is effectively the supremum
of $\lgbrac{\expect{\text{link strength}}}-\expect{\lgbrac{\text{link strength}}}$
over all links and operating regimes of the system. We show that common
fading models including Rayleigh and Nakagami fading have finite logarithmic
Jensen's gap, but some fading models (like bursty fading \cite{wang2013bursty})
have infinite logarithmic Jensen's gap. Subsequently, we show the
usefulness of logarithmic Jensen's gap in obtaining approximate capacity
regions of FF-ICs without instantaneous CSIT. We show that Han-Kobayashi
type rate-splitting schemes \cite{han_kobayashi,chong2008han,etkin_tse_no_fb_IC,suh_tse_fb_gaussian}
based on $inr$, when extended to rate-splitting schemes based on
$\expect{inr}$ for the FF-ICs, give the capacity gap as a function
of logarithmic Jensen's gap, yielding the approximate capacity characterization
for fading models that have finite logarithmic Jensen's gap. Since
our rate-splitting is based on $\expect{inr}$, it does not need instantaneous
CSIT. The constant gap capacity result is first obtained for FF-IC
without feedback or instantaneous CSIT. We also show that for the
FF-IC without feedback, instantaneous CSIT cannot improve the capacity
region over the case with no instantaneous CSIT, except for a constant
gap. We subsequently study FF-IC with feedback and delayed CSIT to
obtain a constant gap capacity result. In this case, having instantaneous
CSIT cannot improve the capacity region over the case with delayed
CSIT. We show that our analysis of FF-IC can easily be extended to
fading interference MAC channel to yield an approximate capacity result.

The usefulness of logarithmic Jensen's gap is further illustrated
by analyzing a scheme based on point-to-point codes for a class of
FF-IC with feedback, where we again obtain capacity gap as a function
of logarithmic Jensen's gap. Our scheme is based on amplify-and-forward
relaying, similar to the one proposed in \cite{suh_tse_fb_gaussian}.
It effectively induces a 2-tap inter-symbol-interference (ISI) channel
for one of the users and a point-to-point feedback channel for the
other user. The work in \cite{suh_tse_fb_gaussian} had similarly
shown that an amplify-and-forward based feedback scheme can achieve
the symmetric rate point, without using rate-splitting. Our scheme
can be considered as an extension to this scheme, which enables us
to approximately achieve the entire capacity region. Our analysis
also yields a capacity bound for a 2-tap fading ISI channel, the tightness
of the bound again determined by the logarithmic Jensen's gap.

The paper is organized as follows. In section \ref{sec:Notation}
we describe the system setup and the notations. In section \ref{sec:fading_models}
we develop the logarithmic Jensen's gap characterization for fading
models. We illustrate a few applications of logarithmic Jensen's gap
characterization in the later sections: in section \ref{sec:FF_IC_nofeedback},
by obtaining approximate capacity region of FF-IC without feedback,
in section \ref{sec:FF_IC_feedback}, by obtaining approximate capacity
region of FF-IC with feedback and delayed CSIT, and in section \ref{sec:FF_IC_p2p_codes},
by developing point-to-point codes for a class of FF-IC with feedback.

\section{Model and Notation\label{sec:Notation}}

\begin{figure}[h]
\begin{centering}
\includegraphics[scale=0.6]{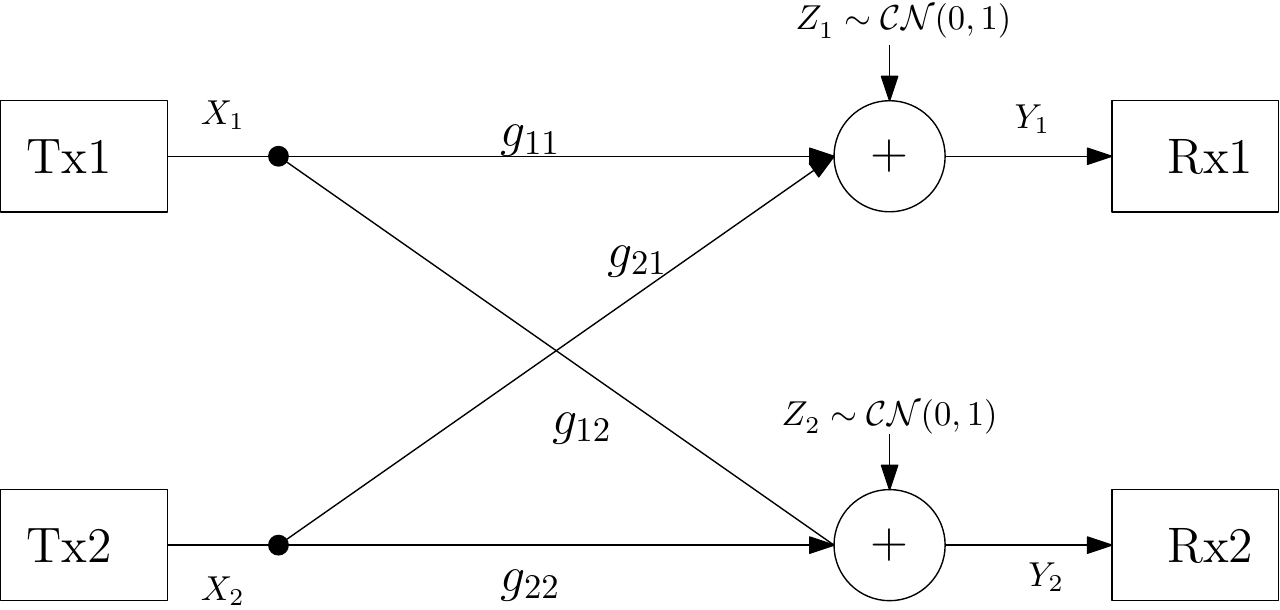}
\par\end{centering}
\caption{\label{fig:nonfeedback_ic}The channel model without feedback.}
\end{figure}

We consider the two-user FF-IC (Figure \ref{fig:nonfeedback_ic})
\begin{align}
Y_{1}(l) & =g_{11}(l)X_{1}(l)+g_{21}(l)X_{2}(l)+Z_{1}(l)\\
Y_{2}(l) & =g_{12}(l)X_{1}(l)+g_{22}(l)X_{2}(l)+Z_{2}(l),
\end{align}
where $Y_{i}(l)$ is the channel output of receiver $i$ (Rx$i$)
at time  $l$, $X_{i}(l)$ is the input of transmitter $i$ (Tx$i$)
at time $l$, $Z_{i}(l)\sim\mathcal{CN}(0,1)$ is complex AWGN noise
process at Rx$i$, and $g_{ij}(l)$ is the time-variant random channel
gain. The channel gain processes $\cbrac{g_{ij}(l)}$ are constant
over a block of size $T$ and independent across blocks and across
links $\brac{i,j}$. Without loss of generality we assume block size
$T=1$ for the fading, our results can be easily extended for arbitrary
$T$ case by coding across the blocks. The transmitters are assumed
to have no knowledge of the channel gain realizations, but the receivers
do have full knowledge of their corresponding channels. %
We assume that $\left|g_{ij}(l)\right|^{2}$ is a random variable
with a known distribution. We assume average power constraint $\frac{1}{N}\sum_{l=1}^{N}\left|X_{i}(l)\right|^{2}\leq1,i=1,2$
at the transmitters, and assume Tx$i$ has a message $W_{i}\in\cbrac{1,\ldotp,2^{NR_{i}}}$,
for a block length of $N$, intended for Rx$i$ for $i=1,2$, and
$W_{1},W_{2}$ are independent of each other. We denote $SNR_{i}:=\expect{\abs{g_{ii}}^{2}}$
for $i=1,2$, and $INR_{i}:=\expect{\abs{g_{ij}}^{2}}$ for $i\neq j$.
For the instantaneous interference channel gains we use $inr_{i}:=\abs{g_{ij}}^{2}$,
$i\neq j$. Note that we allow for arbitrary channel gains, and do
not limit ourselves to SNR-scaling results, but get an approximate
characterization of the FF-IC capacity region.

\begin{figure}[h]
\begin{centering}
\includegraphics[scale=0.6]{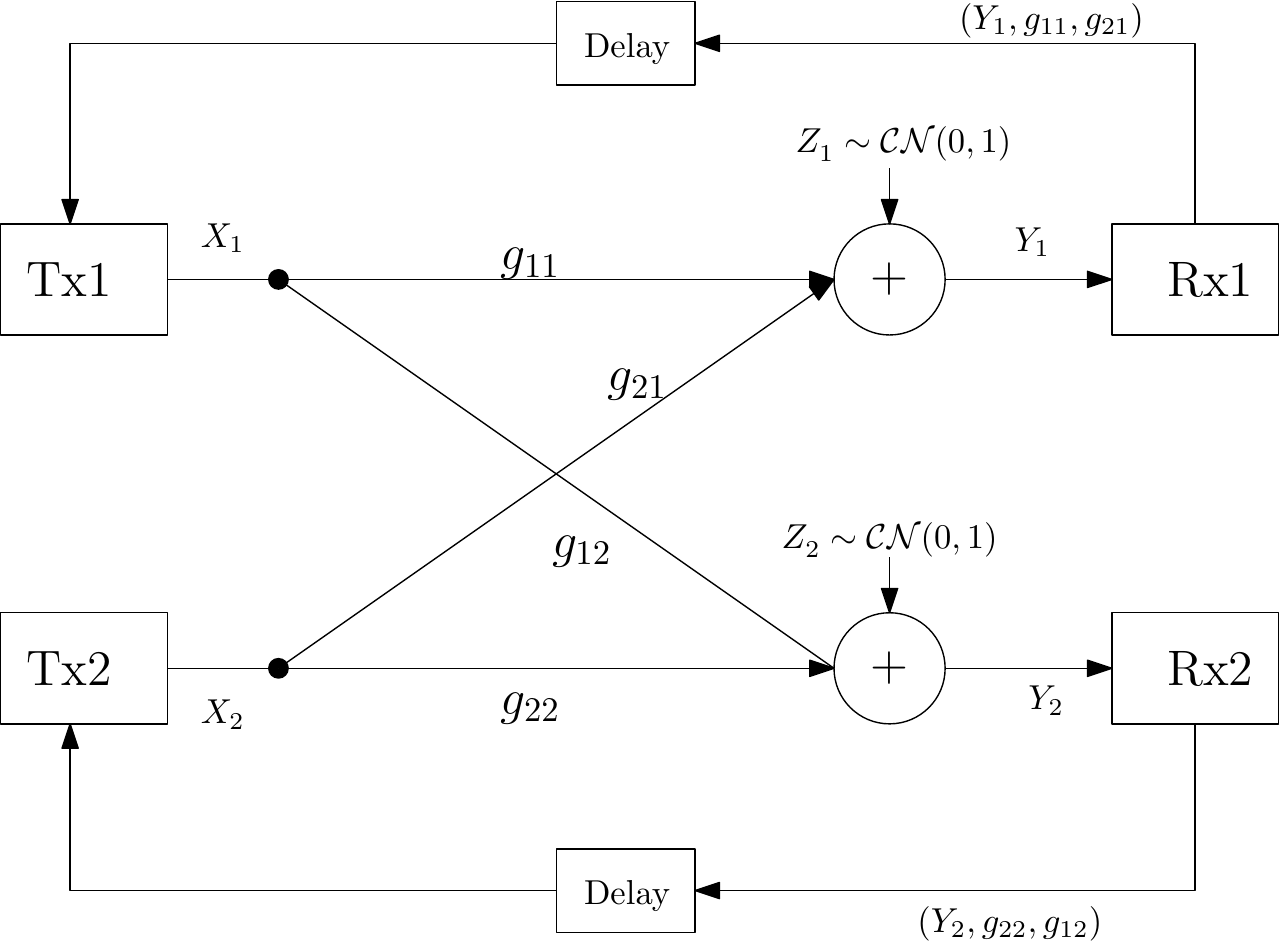}
\par\end{centering}
\caption{\label{fig:feedback_ic}The channel model with feedback.}
\end{figure}

Under the feedback model (Figure \ref{fig:feedback_ic}), after each
reception, each receiver reliably feeds back the received symbol and
the channel states to its corresponding transmitter\footnote{IC with rate limited feedback is considered in \cite{VahidSuh_12_ratelimited}
where outputs are quantized and fed back. Our schemes can also be
extended for such cases.}. For example, at time $l$, Tx$1$ receives $\brac{Y_{1}\brac{l-1},g_{11}(l-1),g_{21}(l-1)}$
from Rx$1$. Thus $X_{1}(l)$ is allowed to be a function of $\left(W_{1},\cbrac{Y_{1}\brac k,g_{11}(k),g_{21}(k)}_{k<l}\right)$.

We define symmetric FF-IC to be a FF-IC such that $g_{11}\sim g_{22}\sim g_{d}$
and $g_{12}\sim g_{21}\sim g_{c}$ (we use the symbol $\sim$ to indicate
random variables following same distribution), all of them being independent.
Here $g_{d}$ and $g_{c}$ are dummy random variables according to
which the direct links and cross links are distributed. We denote
$SNR:=\expect{\abs{g_{d}}^{2}}$, and $INR:=\expect{\abs{g_{c}}^{2}}$,
for the symmetric case.

\begin{figure}[h]
\begin{centering}
\includegraphics[scale=0.75]{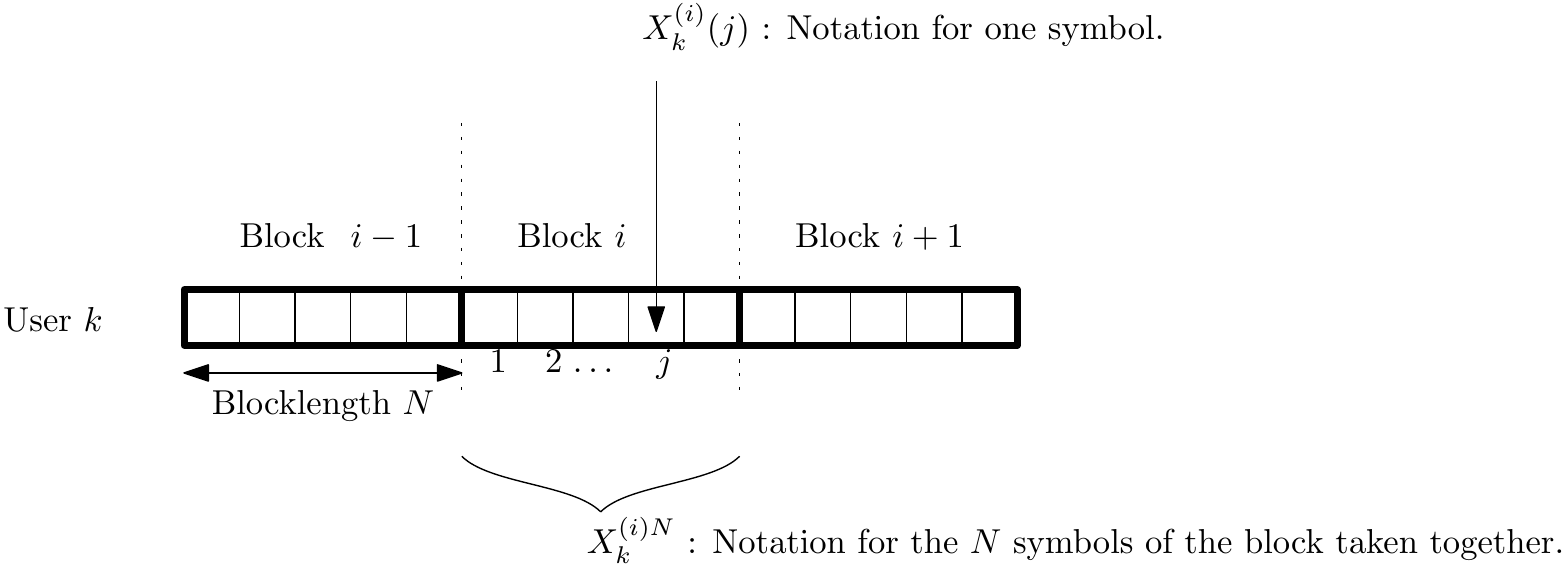}
\par\end{centering}
\caption{\label{fig:block_notation}The notation for schemes involving multiple
blocks (phases).}
\end{figure}

We use the vector notation $\underline{g}_{1}=[g_{11},g_{21}]$, $\underline{g}_{2}=[g_{22},g_{12}]$
and $\underline{g}=[g_{11},g_{21},g_{22},g_{12}]$. For schemes involving
multiple blocks (phases) we use the notation $X_{k}^{(i)N}$, where
$k$ is the user index, $i$ is the block (phase) index and $N$ is
the number of symbols per block. The notation $X_{k}^{\brac i}\brac j$
indicates the $j^{\text{th}}$ symbol in the $i^{\text{th}}$ block
(phase) of $k^{\text{th}}$ user. We explain this in Figure \ref{fig:block_notation}.
{}

The natural logarithm is denoted by $\ln\brac{}$ and the logarithm
with base 2 is denoted by $\lgbrac{}$. Also we define $\log^{+}(\cdot):=\max\left(\log(\cdot),0\right)$.
For obtaining approximate capacity region of ICs, we say that a rate
region $\mathcal{R}$ achieves a capacity gap of $\delta$ if for
any $\brac{R_{1},R_{2}}\in\mathcal{C}$, $\brac{R_{1}-\delta,R_{2}-\delta}\in\mathcal{R}$,
where $\mathcal{C}$ is the capacity region of the channel.

\section{A logarithmic Jensen's gap characterization for fading models\label{sec:fading_models}}

\begin{defn}
\label{def:jensens_gap} For a given fading model, let $\Phi=\cbrac{\phi:\abs{g_{ij}}^{2}\sim\phi,\text{ for some }i,j\in\cbrac{1,2}}$
be the set of all probability density functions, that the fading model
induce on the channel link strengths $\abs{g_{ij}}^{2}$, across all
operating regimes of the system. We define logarithmic Jensen's gap
$c_{JG}$ of the fading model  to be
\begin{equation}
c_{JG}=\sup_{a\in\real^{+},W\sim\phi\in\Phi}\brac{\lgbrac{a+\expect W}-\expect{\lgbrac{a+W}}}.
\end{equation}
\end{defn}
In other words it is the smallest value of $c$ such that
\begin{align}
\lgbrac{a+\expect W}-\expect{\lgbrac{a+W}}\leq c,\label{eq:jensens_gap}
\end{align}
for any $a\geq0$ , for \emph{any} $\phi\in\Phi$, with $W$ distributed
according to $\phi$.

The following lemma converts requirement in Definition~\ref{def:jensens_gap}
to a simpler form.
\begin{lem}
\label{lem:fading_gen} The requirement $\lgbrac{a+\expect W}-\expect{\lgbrac{a+W}}\leq c$
for any $a\geq0$, is equivalent to $\lgbrac{\expect W}-\expect{\lgbrac W}=-\expect{\lgbrac{W'}}\leq c,$
where $W'=\frac{W}{\expect W}$.
\end{lem}
\begin{IEEEproof}
We first note that letting $a=0$ in the requirement $\lgbrac{a+\expect W}-\expect{\lgbrac{a+W}}\leq c$
shows that $\lgbrac{\expect W}-\expect{\lgbrac W}=-\expect{\lgbrac{W'}}\leq c$
is necessary.

To prove the converse, note that $\xi\brac a=\lgbrac{a+\expect W}-\expect{\lgbrac{a+W}}\geq0$
due to Jensen's inequality. Taking derivative with respect to $a$
and again using Jensen's inequality we get
\begin{equation}
\brac{\ln2}\xi'\brac a=\brac{a+\expect W}^{-1}-\expect{\brac{a+W}^{-1}}\leq0.
\end{equation}
Hence, $\xi\brac a$ achieves the maximum value at $a=0$ in the range
$[0,\infty)$. Hence we have the equivalent condition
\begin{equation}
\lgbrac{\expect W}-\expect{\lgbrac W}\leq c,
\end{equation}
which is equivalent to $-\expect{\lgbrac{W'}}\leq c.$
\end{IEEEproof}
Hence, it follows that for any distribution that has a point mass
at $0$ (for example, bursty interference model \cite{wang2013bursty}),
we do not have a finite logarithmic Jensen's gap, since it has $\expect{\lgbrac{W'}}=-\infty$.
Now we discuss a few distributions that can be easily shown to have
a finite logarithmic Jensen's gap. Note that any finite $c$, which
satisfies Equation \eqref{eq:jensens_gap}, is an upper bound to the
logarithmic Jensen's gap $c_{JG}$.%

\subsection{Gamma distribution}

Gamma distribution generalizes some of the commonly used fading models,
including Rayleigh and Nakagami fading. The probability density function
for Gamma distribution is given by
\begin{equation}
f\brac w=\frac{w^{k-1}e^{-\frac{w}{\theta}}}{\theta^{k}\Gamma(k)}
\end{equation}
for $w>0$, where $k>0$ is the shape parameter, and $\theta>0$ is
the scale parameter.
\begin{prop}
If the elements of $\Phi$ are Gamma distributed with shape parameter
$k$, they satisfy Equation \eqref{eq:jensens_gap} with constant
$c=\frac{\lgbrac e}{k}-\lgbrac{1+\frac{1}{2k}}$.
\end{prop}
\begin{IEEEproof}
Using Lemma \ref{lem:fading_gen}, it is sufficient to prove $\lgbrac{\expect W}-\expect{\lgbrac W}\leq\frac{\lgbrac e}{\alpha}-\lgbrac{1+\frac{1}{2\alpha}}$.
It is known for the Gamma distribution that $\expect W=k\theta$ and
$\expect{\lnbrac W}=\psi\brac k+\lnbrac{\theta},$ where $\psi$ is
the digamma function. Therefore,
\begin{equation}
\begin{aligned}\lgbrac{\expect W}-\expect{\lgbrac W} & =\log(e)\left(\lnbrac k-\psi\brac k\right)\end{aligned}
.
\end{equation}
We first use the following property of digamma function
\begin{equation}
\psi\brac k=\psi\brac{k+1}-\frac{1}{k},
\end{equation}
and then use the inequality from \cite[Lemma 1.7]{batir2008gammainequalities}
\begin{equation}
\lnbrac{k+\frac{1}{2}}<\psi\brac{k+1}.
\end{equation}
Hence,
\begin{align}
\lgbrac{\expect W}-\expect{\lgbrac W} & <\log(e)\left(\lnbrac k-\lnbrac{k+\frac{1}{2}}+\frac{1}{k}\right)\nonumber \\
 & =\frac{\lgbrac e}{k}-\lgbrac{1+\frac{1}{2k}}.\label{eq:gamma_c}
\end{align}
\end{IEEEproof}
\begin{cor}
If the elements of $\Phi$ are exponentially distributed (which corresponds
to Rayleigh fading), they satisfy Equation \eqref{eq:jensens_gap}
with constant $c=0.86$.
\end{cor}
\begin{IEEEproof}
In Rayleigh fading model the $\abs{g_{ij}}^{2}$ is exponentially
distributed. The exponential distribution itself is a special case
of Gamma distribution with $k=1$ . Substituting $\alpha=1$ in $(\ref{eq:gamma_c})$
we get $\lgbrac{\expect W}-\expect{\lgbrac W}<0.86$.
\end{IEEEproof}
Nakagami fading can be obtained as a special case of the Gamma distribution;
in this case the logarithmic Jensen's gap will depend upon the parameters
used in the model.

\subsection{Weibull distribution}

The probability density function for Weibull distribution is given
by
\begin{equation}
f\brac w=\frac{k}{\lambda}\left(\frac{w}{\lambda}\right)^{k-1}e^{-(w/\lambda)^{k}}
\end{equation}
for $x>0$ with $k,\lambda>0$.
\begin{prop}
If the elements of $\Phi$ are Weibull distributed with parameter
$k$, they satisfy Equation \eqref{eq:jensens_gap} with $c=\frac{\gamma\lgbrac e}{k}+\lgbrac{\Gamma\brac{1+\frac{1}{k}}}$,
where $\gamma$ is Euler's constant.
\end{prop}
\begin{IEEEproof}
For Weibull distributed $W$, we have $\expect W=\lambda\Gamma\brac{1+\frac{1}{k}}$
and $\expect{\lnbrac W}=\lnbrac{\lambda}-\frac{\gamma}{k}$, where
$\Gamma\brac{\cdot}$ denotes the gamma function and $\gamma$ is
the Euler's constant. Hence, it follows that
\begin{equation}
\lgbrac{\expect W}-\expect{\lgbrac W}\leq\frac{\gamma\lgbrac e}{k}+\lgbrac{\Gamma\brac{1+\frac{1}{k}}}.
\end{equation}
Using Lemma $\ref{lem:fading_gen}$ concludes the proof.
\end{IEEEproof}
Note that exponential distribution can be specialized from Weibull
distribution as well, by setting $k=1$. Hence, we get the tighter
gap in the following corollary.
\begin{cor}
If the elements of $\Phi$ are exponentially distributed, they satisfy
Equation \eqref{eq:jensens_gap} with constant $c=0.83$.
\end{cor}
In the following table we summarize the values we obtain as upper
bound on logarithmic Jensen's gap, according to Definition \ref{def:jensens_gap}
and Equation \eqref{eq:jensens_gap} for different fading models.

\begin{table}[htbp]
\centering{}\caption{\label{tab:gap_for_distributions} Upper bound of logarithmic Jensen's
gap for different fading models}
\begin{tabular}{|c|c|}
\hline
Fading Model  & $c$ \tabularnewline
\hline
\hline
Rayleigh & $0.83$\tabularnewline
\hline
Gamma $k=1$ & $0.86$\tabularnewline
\hline
Gamma $k=2$ & $0.40$\tabularnewline
\hline
Gamma $k=3$ & $0.26$\tabularnewline
\hline
Weibull $k=1$ & $0.83$\tabularnewline
\hline
Weibull $k=2$ & $0.24$\tabularnewline
\hline
Weibull $k=3$ & $0.11$\tabularnewline
\hline
\end{tabular}
\end{table}

\subsection{Other distributions}

Here we give a lemma that can be used together with Lemma \ref{lem:fading_gen}
to verify whether a given fading model has a finite logarithmic Jensen's
gap.%

\begin{lem}
If the cumulative distribution function $F\brac w$ of $W$ satisfies
$F\brac w\leq aw^{b}$ over $w\in[0,\epsilon]$ for some $a\geq0$,
$b>0$, and $0<\epsilon\leq1$, then
\begin{equation}
\expect{\lnbrac W}\geq\lnbrac{\epsilon}+a\epsilon^{b}\lnbrac{\epsilon}-\frac{a\epsilon^{b}}{b}.
\end{equation}
\label{lem:other_distr}
\end{lem}
\begin{IEEEproof}
The condition in this lemma ensures that the probability density function
$f(w)$ grows slow enough as $w\rightarrow0^{-}$ so that $f(w)\lnbrac w$
is integrable at $0$. Also the behavior for large values of $w$
is not relevant here, since we are looking for a lower bound on $\expect{\lnbrac W}$.
The detailed proof is in \ifarxiv Appendix \ref{app:proof_bounded_log_lemma}\else  \cite[Appendix C]{Joyson_jensens_gap_arxiv}\fi.
\end{IEEEproof}
Hence, if the cumulative distribution of the channel gain grows polynomially
in a neighborhood of $0$, the resulting logarithm becomes integrable,
and thus it is possible to find a finite constant $c$ for the Equation
\eqref{eq:jensens_gap}.

\section{Approximate Capacity Region of FF-IC without feedback\label{sec:FF_IC_nofeedback}}

In this section we make use of the logarithmic Jensen's gap characterization
to obtain the approximate capacity region of FF-IC with neither feedback
nor instantaneous CSIT.%

\begin{thm}
\label{th:splitting-nfb} For a non-feedback FF-IC with a finite logarithmic
Jensen's gap $c_{JG}$ , the rate region $\mathcal{R}_{NFB}$ described
by \eqref{eq:inner_nofb} is achievable with $\lambda_{pk}=\min\brac{\frac{1}{INR_{k}},1}$:
\begin{subequations}\label{eq:inner_nofb}
\begin{align}
R_{1} & \leq\expect{\lgbrac{1+\abs{g_{11}}^{2}+\lambda_{p2}\abs{g_{21}}^{2}}}-1\label{eq:inner_nofb1}\\
R_{2} & \leq\expect{\lgbrac{1+\abs{g_{22}}^{2}+\lambda_{p1}\abs{g_{12}}^{2}}}-1\label{eq:inner_nofb2}\\
R_{1}+R_{2} & \leq\expect{\lgbrac{1+\abs{g_{22}}^{2}+\abs{g_{12}}^{2}}}+\expect{\lgbrac{1+\lambda_{p1}\abs{g_{11}}^{2}+\lambda_{p2}\abs{g_{21}}^{2}}}-2\label{eq:inner_nofb3}\\
R_{1}+R_{2} & \leq\expect{\lgbrac{1+\abs{g_{11}}^{2}+\abs{g_{21}}^{2}}}+\expect{\lgbrac{1+\lambda_{p2}\abs{g_{22}}^{2}+\lambda_{p1}\abs{g_{12}}^{2}}}-2\label{eq:inner_nofb4}\\
R_{1}+R_{2} & \leq\expect{\lgbrac{1+\lambda_{p1}\abs{g_{11}}^{2}+\abs{g_{21}}^{2}}}+\expect{\lgbrac{1+\lambda_{p2}\abs{g_{22}}^{2}+\abs{g_{12}}^{2}}}-2\label{eq:inner_nofb5}\\
2R_{1}+R_{2} & \leq\expect{\lgbrac{1+\abs{g_{11}}^{2}+\abs{g_{21}}^{2}}}+\expect{\lgbrac{1+\lambda_{p2}\abs{g_{22}}^{2}+\abs{g_{12}}^{2}}}\nonumber \\
 & \qquad+\expect{\lgbrac{1+\lambda_{p1}\abs{g_{11}}^{2}+\lambda_{p2}\abs{g_{21}}^{2}}}-3\label{eq:inner_nofb6}\\
R_{1}+2R_{2} & \leq\expect{\lgbrac{1+\abs{g_{22}}^{2}+\abs{g_{12}}^{2}}}+\expect{\lgbrac{1+\lambda_{p1}\abs{g_{11}}^{2}+\abs{g_{21}}^{2}}}\nonumber \\
 & \qquad+\expect{\lgbrac{1+\lambda_{p2}\abs{g_{22}}^{2}+\lambda_{p1}\abs{g_{12}}^{2}}}-3\label{eq:inner_nofb7}
\end{align}
\end{subequations}and the region $\mathcal{R}_{NFB}$ has a capacity
gap of at most $c_{JG}+1$ bits per channel use.
\end{thm}
\begin{IEEEproof}
This region is obtained by a rate-splitting scheme that allocates
the private message power proportional to $\frac{1}{\expect{inr}}.$
The analysis of the scheme and outer bounds are similar to that in
\cite{etkin_tse_no_fb_IC}. See subsection \ref{subsec:no_fb} for
details.
\end{IEEEproof}
\begin{rem}
For the case of Rayleigh fading we obtain a capacity gap of $1.83$
bits per channel use, following Table \ref{tab:gap_for_distributions}.
\end{rem}

\begin{cor}
For FF-IC with finite logarithmic Jensen's gap $c_{JG}$, instantaneous
CSIT cannot improve the capacity region of except by a constant.
\end{cor}
\begin{IEEEproof}
Our outer bounds in subsection \ref{subsec:no_fb} for the non-feedback
IC are valid even when there is instantaneous CSIT. These outer bounds
are within constant gap of the rate region $\mathcal{R}_{NFB}$ achieved
without instantaneous CSIT.
\end{IEEEproof}
\begin{cor}
Delayed CSIT cannot improve the capacity region of the FF-IC except
by a constant.
\end{cor}
\begin{IEEEproof}
This follows from the previous corollary, since instantaneous CSIT
is always better than delayed CSIT.
\end{IEEEproof}
\begin{rem}
The previous two corollaries are for FF-IC with 2 users and single
antennas. It does not contradict the results for MISO broadcast channel,
X-channel, MIMO IC and multi-user IC where delayed CSIT or instantaneous
CSIT can improve capacity region by more than a constant \cite{retrospective_maddah_ali_tse,retrospective_interfer_align_Maleki_2012,ergodic_delayed_kang2013,ergodic_interference_nazer2012,MIMO_IC_delayed_csit_Vaze_2012}.
\end{rem}
\begin{cor}
Within a constant gap, the capacity region of the FF-IC (with finite
logarithmic Jensen's gap $c_{JG}$) can be proved to be same as the
capacity region of IC (without fading) with equivalent channel strengths
$SNR_{i}:=\expect{\abs{g_{ii}}^{2}}$ for $i=1,2$, and $INR_{i}:=\expect{\abs{g_{ij}}^{2}}$
for $i\neq j$.\label{cor:fading=00003Dstatic_nofb}
\end{cor}
\begin{IEEEproof}
This is an application of the logarithmic Jensen's gap result. The
proof is given in  \ifarxiv Appendix \ref{app:proof_cor_fading_is_static_nofb}\else  \cite[Appendix D]{Joyson_jensens_gap_arxiv}\fi.
\end{IEEEproof}

\subsection{Discussion}

It is useful to view Theorem \ref{th:splitting-nfb} in the context
of the existing results for the ICs. It is known that for ICs, one
can approximately achieve the capacity region by performing superposition
coding and allocating a power to the private symbols that is inversely
proportional to the strength of the interference caused at the unintended
receiver. Consequently, the received interference power is at the
noise level, and the private symbols can be safely treated as noise,
incurring only a constant rate penalty. At first sight, such a strategy
seems impossible for the fading IC, where the transmitters do not
have instantaneous channel information. What Theorem \ref{th:splitting-nfb}
reveals (with the details in subsection \ref{subsec:no_fb}) is that
if the fading model has finite logarithmic Jensen's gap, it is sufficient
to perform power allocation based on the inverse of average interference
strength to approximately achieve the capacity region.

\begin{figure}
\begin{centering}
\includegraphics[scale=0.65]{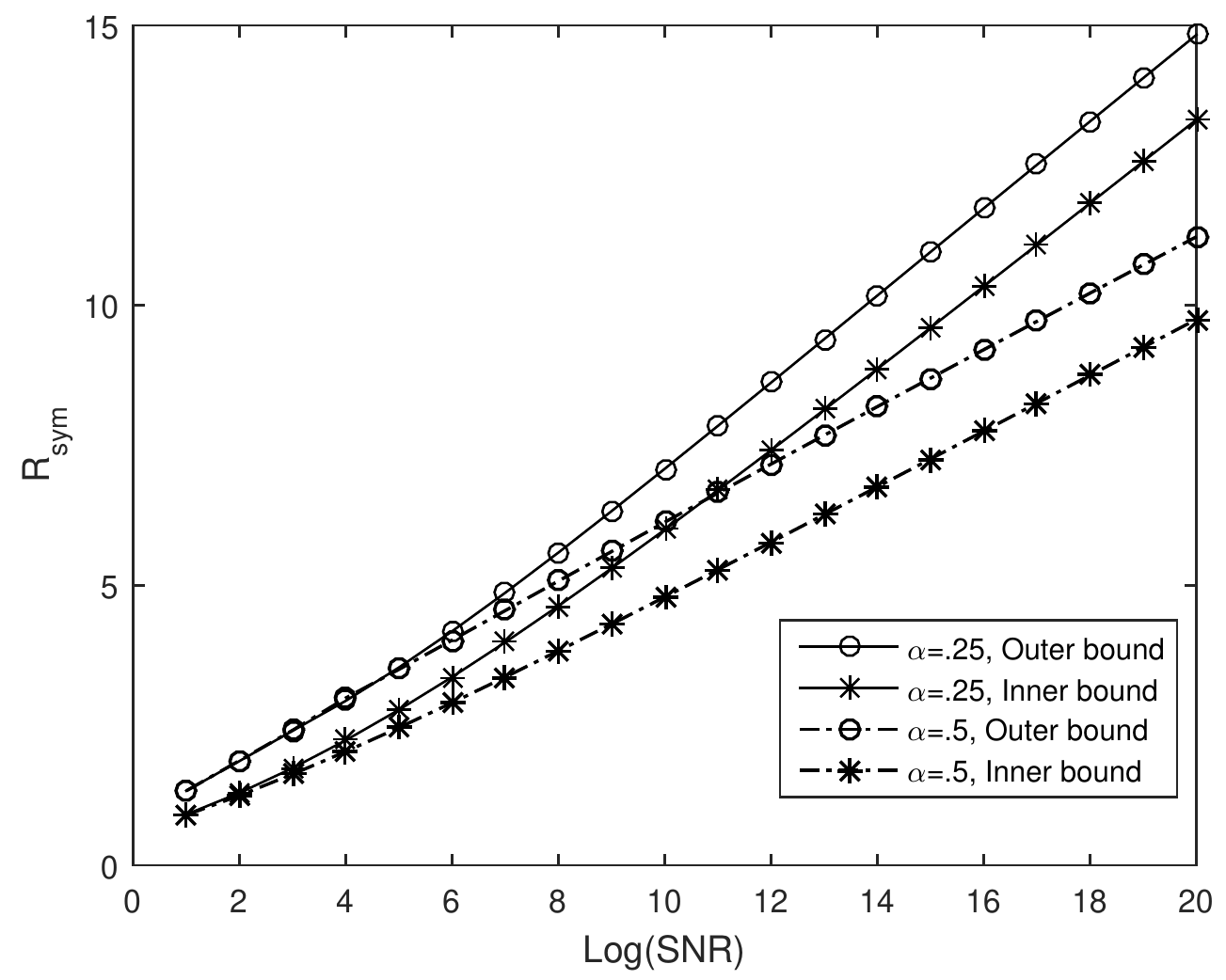}
\par\end{centering}
\caption{\label{fig:nfb_alpha_0p5}Comparison of outer and inner bounds with
given $\alpha=\frac{\protect\lgbrac{INR}}{\protect\lgbrac{SNR}}$
for non-feedback symmetric FF-IC at the symmetric rate point. For
high SNR, the capacity gap is approximately $1.48$ bits per channel
use for $\alpha=0.5$ and $1.51$ bits per channel use for $\alpha=0.25$
from the numerics. Our theoretical analysis yields gap as $1.83$
bits per channel use independent of $\alpha$.}
\end{figure}

We compare the symmetric rate point achievable for the non-feedback
symmetric FF-IC in Figure \ref{fig:nfb_alpha_0p5}. The fading model
used is Rayleigh fading. The inner bound in numerical simulation is
from Equation \eqref{eq:ach_nofb} (which is slightly tighter than
\eqref{eq:inner_nofb} since some terms in \eqref{eq:ach_nofb} are
simplified and bounded with the worst case values to obtain \eqref{eq:inner_nofb})
in subsection \ref{subsec:no_fb} according to the choice of distributions
given in the same subsection. The outer bound is plotted by simulating
Equation \eqref{eq:outer_nofb} in subsection \ref{subsec:no_fb}.
The $SNR$ is varied after fixing $\alpha=\frac{\lgbrac{INR}}{\lgbrac{SNR}}$.
The simulation yields a capacity gap of $1.48$ bits per channel use
for $\alpha=0.5$ and a capacity gap of $1.51$ bits per channel use
for $\alpha=0.25$. Our theoretical analysis for FF-IC gives a capacity
gap of $c_{JG}+1\leq1.83$ bits per channel use independent of $\alpha$,
using data from Table \ref{tab:gap_for_distributions} in Section
\ref{sec:fading_models}.

\subsection{Proof of Theorem~\ref{th:splitting-nfb}\label{subsec:no_fb}}

From \cite{chong2008han} we obtain that a Han-Kobayashi scheme for
IC can achieve the following rate region for all $p\brac{u_{1}}p\brac{u_{2}}p\brac{x_{1}|u_{1}}p\brac{x_{2}|u_{2}}$.
Note that we use $\brac{Y_{i},\underline{g_{i}}}$ instead of $\brac{Y_{i}}$
in the actual result from \cite{chong2008han} to account for the
fading. \begin{subequations}\label{eq:ach_nofb}
\begin{align}
R_{1} & \leq I\brac{X_{1};Y_{1},\underline{g_{1}}|U_{2}}\label{eq:ach_nofb1}\\
R_{2} & \leq I\brac{X_{2};Y_{2},\underline{g_{2}}|U_{1}}\label{eq:ach_nofb2}\\
R_{1}+R_{2} & \leq I\brac{X_{2},U_{1};Y_{2},\underline{g_{2}}}+I\brac{X_{1};Y_{1},\underline{g_{1}}|U_{1},U_{2}}\label{eq:ach_nofb3}\\
R_{1}+R_{2} & \leq I\brac{X_{1},U_{2};Y_{1},\underline{g_{1}}}+I\brac{X_{2};Y_{2},\underline{g_{2}}|U_{1},U_{2}}\label{eq:ach_nofb4}\\
R_{1}+R_{2} & \leq I\brac{X_{1},U_{2};Y_{1},\underline{g_{1}}|U_{1}}+I\brac{X_{2},U_{1};Y_{2},\underline{g_{2}}|U_{2}}\label{eq:ach_nofb5}\\
2R_{1}+R_{2} & \leq I\brac{X_{1},U_{2};Y_{1},\underline{g_{1}}}+I\brac{X_{1};Y_{1},\underline{g_{1}}|U_{1},U_{2}}+I\brac{X_{2},U_{1};Y_{2},\underline{g_{2}}|U_{2}}\label{eq:ach_nofb6}\\
R_{1}+2R_{2} & \leq I\brac{X_{2},U_{1};Y_{2},\underline{g_{2}}}+I\brac{X_{2};Y_{2},\underline{g_{2}}|U_{1},U_{2}}+I\brac{X_{1},U_{2};Y_{1},\underline{g_{1}}|U_{1}}.\label{eq:ach_nofb7}
\end{align}
\end{subequations}Now similar to that in \cite{etkin_tse_no_fb_IC},
choose mutually independent Gaussian input distributions $U_{k},X_{pk}$
to generate $X_{k}$.
\begin{equation}
U_{k}\sim\mathcal{CN}\brac{0,\lambda_{ck}},\quad X_{pk}\sim\mathcal{CN}\brac{0,\lambda_{pk}},\qquad k\in\cbrac{1,2}
\end{equation}
\begin{equation}
X_{1}=U_{1}+X_{p1},\quad X_{2}=U_{2}+X_{p2},
\end{equation}
where $\lambda_{ck}+\lambda_{pk}=1$ and $\lambda_{pk}=\min\brac{\frac{1}{INR_{k}},1}$.
Here we introduced the rate-splitting using the average $inr$. On
evaluating the region described by \eqref{eq:ach_nofb} with this
choice of input distribution, we get the region described by \eqref{eq:inner_nofb};
the calculations are deferred to Appendix \ref{app:achievability_no_fb}.
\begin{claim}
An outer bound for the non-feedback case is given by \eqref{eq:outer_nofb}:
\begin{subequations}\label{eq:outer_nofb}
\begin{align}
R_{1} & \leq\expect{\lgbrac{1+\abs{g_{11}}^{2}}}\label{eq:outer_nofb1}\\
R_{2} & \leq\expect{\lgbrac{1+\abs{g_{22}}^{2}}}\label{eq:outer_nofb2}\\
R_{1}+R_{2} & \leq\expect{\lgbrac{1+\abs{g_{22}}^{2}+\abs{g_{12}}^{2}}}+\expect{\lgbrac{1+\frac{\abs{g_{11}}^{2}}{1+\abs{g_{12}}^{2}}}}\label{eq:outer_nofb3}\\
R_{1}+R_{2} & \leq\expect{\lgbrac{1+\abs{g_{11}}^{2}+\abs{g_{21}}^{2}}}+\expect{\lgbrac{1+\frac{\abs{g_{22}}^{2}}{1+\abs{g_{21}}^{2}}}}\label{eq:outer_nofb4}\\
R_{1}+R_{2} & \leq\expect{\lgbrac{1+\abs{g_{21}}^{2}+\frac{\abs{g_{11}}^{2}}{1+\abs{g_{12}}^{2}}}}+\expect{\lgbrac{1+\abs{g_{12}}^{2}+\frac{\abs{g_{22}}^{2}}{1+\abs{g_{21}}^{2}}}}\label{eq:outer_nofb5}\\
2R_{1}+R_{2} & \leq\expect{\lgbrac{1+\abs{g_{11}}^{2}+\abs{g_{21}}^{2}}}+\expect{\lgbrac{1+\abs{g_{12}}^{2}+\frac{\abs{g_{22}}^{2}}{1+\abs{g_{21}}^{2}}}}\nonumber \\
 & \qquad+\expect{\lgbrac{1+\frac{\abs{g_{11}}^{2}}{1+\abs{g_{12}}^{2}}}}\label{eq:outer_nofb6}\\
R_{1}+2R_{2} & \leq\expect{\lgbrac{1+\abs{g_{22}}^{2}+\abs{g_{12}}^{2}}}+\expect{\lgbrac{1+\abs{g_{21}}^{2}+\frac{\abs{g_{11}}^{2}}{1+\abs{g_{12}}^{2}}}}\nonumber \\
 & \qquad+\expect{\lgbrac{1+\frac{\abs{g_{22}}^{2}}{1+\abs{g_{21}}^{2}}}}.\label{eq:outer_nofb7}
\end{align}
\end{subequations}
\end{claim}
\begin{IEEEproof}
The outer bounds \eqref{eq:outer_nofb1} and \eqref{eq:outer_nofb2}
are easily derived by removing the interference from the other user
by providing it as side-information.%

The outer bound in Equation \eqref{eq:outer_nofb5} follows from \cite[Theorem 1]{etkin_tse_no_fb_IC}.
Those in Equation \eqref{eq:outer_nofb6} and Equation \eqref{eq:outer_nofb7}
follow from \cite[Theorem 4]{etkin_tse_no_fb_IC}. We just need to
modify the theorems from \cite{etkin_tse_no_fb_IC} for the fading
case by treating $\left(Y_{i},\underline{g}_{i}\right)$ as output,
and using the i.i.d property of the channels. We illustrate the procedure
for Equation \eqref{eq:outer_nofb7} in Appendix \ref{app:outerbounds_nofb}.
Equation \eqref{eq:outer_nofb5} and Equation \eqref{eq:outer_nofb6}
can be derived similarly.

The derivation of outer bounds \eqref{eq:outer_nofb3} and \eqref{eq:outer_nofb4}
uses similar techniques as for Equation \eqref{eq:outer_nofb7}. We
derive Equation \eqref{eq:outer_nofb4} in Appendix \ref{app:outerbounds_nofb}.
Equation \eqref{eq:outer_nofb3} follows due to symmetry.
\end{IEEEproof}
\begin{claim}
The gap between the inner bound \eqref{eq:inner_nofb} and the outer
bound \eqref{eq:outer_nofb} for the non-feedback case is at most
$c_{JG}+1$ bits per channel use.\label{claim:capacity_gap_noFB}
\end{claim}
\begin{IEEEproof}
The proof for the capacity gap uses the logarithmic Jensen's gap property
of the fading model. Denote the gap between the first outer bound
$(\ref{eq:outer_nofb1})$ and first inner bound $(\ref{eq:inner_nofb1})$
by $\delta_{1}$, for the second pair denote the gap by $\delta_{2}$,
and so on. By inspection $\delta_{1}\leq1$ and $\delta_{2}\leq1$.
Now
\begin{align}
\delta_{3} & =\expect{\lgbrac{1+\frac{\abs{g_{11}}^{2}}{1+\abs{g_{12}}^{2}}}}-\expect{\lgbrac{1+\lambda_{p1}\abs{g_{11}}^{2}+\lambda_{p2}\abs{g_{21}}^{2}}}+2\\
 & \overset{(a)}{\leq}\expect{\lgbrac{1+\frac{\abs{g_{11}}^{2}}{1+INR_{1}}}}-\expect{\lgbrac{1+\lambda_{p1}\abs{g_{11}}^{2}}}+2+c_{JG}\\
 & \overset{(b)}{\leq}2+c_{JG}.
\end{align}
The step $(a)$ follows from Jensen's inequality and logarithmic Jensen's
gap property of $\abs{g_{12}}^{2}$. The step $\brac b$ follows because
$\lambda_{p1}=\min\brac{\frac{1}{INR_{1}},1}\geq\frac{1}{INR_{1}+1}$.
Similarly, we can bound the other $\delta$'s and gather the inequalities
as:
\begin{equation}
\delta_{1},\delta_{2}\leq1;\quad\delta_{3},\delta_{4}\leq2+c_{JG};\quad\delta_{5}\leq2+2c_{JG};\quad\delta_{6},\delta_{7}\leq3+2c_{JG}.
\end{equation}
For $\delta_{5},\delta_{6},$ and $\delta_{7}$ we have to use the
logarithmic Jensen's gap property twice and hence $2c_{JG}$ appears.
We note that $\delta_{1}$ is associated with bounding $R_{1}$, $\delta_{2}$
with $R_{2}$, $\brac{\delta_{3},\delta_{4},\delta_{5}}$ with $R_{1}+R_{2}$,
$\delta_{6}$ with $2R_{1}+R_{2}$ and $\delta_{7}$ with $R_{1}+2R_{2}$.
Hence, it follows that the capacity gap is at most $\max\brac{\delta_{1},\delta_{2},\frac{\delta_{3}}{2},\frac{\delta_{4}}{2},\frac{\delta_{5}}{2},\frac{\delta_{6}}{3},\frac{\delta_{7}}{3}}\leq c_{JG}+1$
bits per channel use.
\end{IEEEproof}

\subsection{Fast Fading Interference MAC channel}

\begin{figure}
\begin{centering}
\includegraphics[scale=0.6]{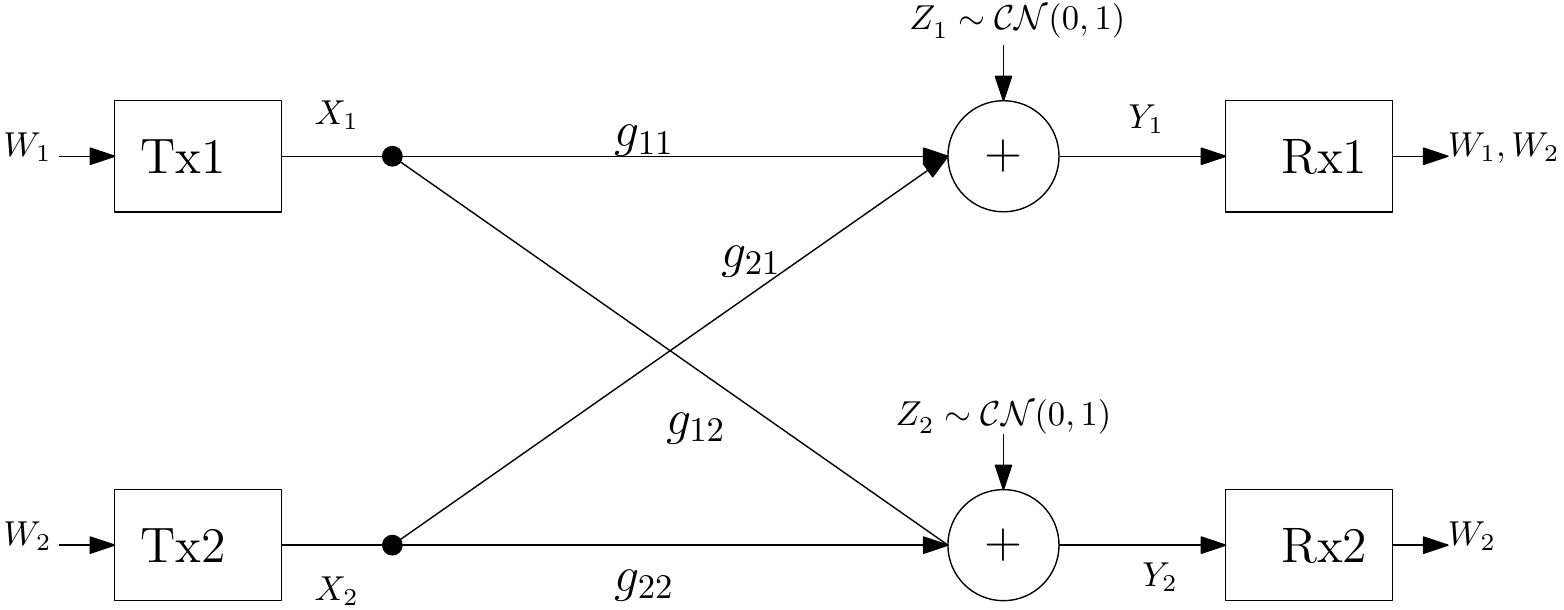}
\par\end{centering}
\caption{Fast Fading Interference MAC channel\label{fig:IMA}}
\end{figure}

We now consider the interference MAC channel \cite{perron_IMA} with
fading links (Figure \ref{fig:IMA}), where we can obtain an approximate
capacity result similar to the FF-IC. This setup has similar network
structure as FF-IC. However Rx1 needs to decode the messages from
both the two transmitters, while Rx2 needs to decode only the message
from Tx2.

\begin{thm}
A rate splitting scheme based on average INR can achieve the approximate
capacity region of fast fading interference MAC channel with a finite
logarithmic Jensen's gap $c_{JG}$, within $1+\frac{1}{2}c_{JG}$
bits per channel use.\label{thm:IMA}
\end{thm}
\begin{IEEEproof}
The proof is by extending the techniques used in \cite{perron_IMA}
and using similar calculations as for the FF-IC. Details are in \ifarxiv  Appendix \ref{app:IMA}\else  \cite[Appendix E]{Joyson_jensens_gap_arxiv}\fi.
\end{IEEEproof}

\section{Approximate Capacity Region of FF-IC with feedback \label{sec:FF_IC_feedback}}

In this section we make use of the logarithmic Jensen's gap characterization
to obtain the approximate capacity region of FF-IC with output and
channel state feedback, but transmitters having no prior knowledge
of channel states. Under the feedback model, after each reception,
each receiver reliably feeds back the received symbol and the channel
states to its corresponding transmitter. For example, at time $l$,
Tx$1$ receives $\brac{Y_{1}\brac{l-1},g_{11}(l-1),g_{21}(l-1)}$
from Rx$1$. Thus $X_{1}(l)$ is allowed to be a function of $\left(W_{1},\cbrac{Y_{1}\brac k,g_{11}(k),g_{21}(k)}_{k<l}\right)$.
The model is described in section \ref{sec:Notation} and is illustrated
with Figure \ref{fig:feedback_ic} in the same section.
\begin{thm}
\label{th:splitting-fb} For a feedback FF-IC with a finite logarithmic
Jensen's gap $c_{JG}$ , the rate region $\mathcal{R}_{FB}$ described
by \eqref{eq:inner_fb} is achievable for $0\leq\abs{\rho}^{2}\leq1$,
$0\leq\theta<2\pi$ with $\lambda_{pk}=\min\brac{\frac{1}{INR_{k}},1-\abs{\rho}^{2}}$:\label{thm:apprx_capacity_fb}
\begin{subequations}\label{eq:inner_fb}
\begin{align}
R_{1} & \leq\expect{\lgbrac{\abs{g_{11}}^{2}+\abs{g_{21}}^{2}+2\abs{\rho}^{2}\text{Re}\brac{e^{i\theta}g_{11}g_{21}^{*}}+1}}-1\label{eq:inner_fb1}\\
R_{1} & \leq\expect{\lgbrac{1+\brac{1-\abs{\rho}^{2}}\abs{g_{12}}^{2}}}+\expect{\lgbrac{1+\lambda_{p1}\abs{g_{11}}^{2}+\lambda_{p2}\abs{g_{21}}^{2}}}-2\label{eq:inner_fb2}\\
R_{2} & \leq\expect{\lgbrac{\abs{g_{22}}^{2}+\abs{g_{12}}^{2}+2\abs{\rho}^{2}\text{Re}\brac{e^{i\theta}g_{22}^{*}g_{12}}+1}}-1\label{eq:inner_fb3}\\
R_{2} & \leq\expect{\lgbrac{1+\brac{1-\abs{\rho}^{2}}\abs{g_{21}}^{2}}}+\expect{\lgbrac{1+\lambda_{p2}\abs{g_{22}}^{2}+\lambda_{p1}\abs{g_{12}}^{2}}}-2\label{eq:inner_fb4}\\
R_{1}+R_{2} & \leq\expect{\lgbrac{\abs{g_{22}}^{2}+\abs{g_{12}}^{2}+2\abs{\rho}^{2}\text{Re}\brac{e^{i\theta}g_{22}^{*}g_{12}}+1}}\nonumber \\
 & \qquad+\expect{\lgbrac{1+\lambda_{p1}\abs{g_{11}}^{2}+\lambda_{p2}\abs{g_{21}}^{2}}}-2\label{eq:inner_fb5}\\
R_{1}+R_{2} & \leq\expect{\lgbrac{\abs{g_{11}}^{2}+\abs{g_{21}}^{2}+2\abs{\rho}^{2}\text{Re}\brac{e^{i\theta}g_{11}g_{21}^{*}}+1}}\nonumber \\
 & \qquad+\expect{\lgbrac{1+\lambda_{p2}\abs{g_{22}}^{2}+\lambda_{p1}\abs{g_{12}}^{2}}}-2\label{eq:inner_fb6}
\end{align}
\end{subequations}and the region $\mathcal{R}_{FB}$ has a capacity
gap of at most $c_{JG}+2$ bits per channel use.
\end{thm}
\begin{IEEEproof}
The proof is in subsection \ref{subsec:fb}.
\end{IEEEproof}

\begin{rem}
For the case of Rayleigh fading we obtain a capacity gap of $2.83$
bits per channel use, following Table \ref{tab:gap_for_distributions}.
\end{rem}
\begin{cor}
Instantaneous CSIT cannot improve the capacity region of the FF-IC
(with finite logarithmic Jensen's gap $c_{JG}$) with feedback and
delayed CSIT except for a constant.
\end{cor}
\begin{IEEEproof}
Our outer bounds in subsection \ref{subsec:fb} for feedback case
are valid even when there is instantaneous CSIT. These outer bounds
are within constant gap of the rate region $\mathcal{R}_{FB}$ achieved
using only feedback and delayed CSIT.
\end{IEEEproof}
\begin{cor}
If the phases of the links $g_{ij}$ are uniformly distributed in
$\sbrac{0,2\pi}$, then within a constant gap, the capacity region
of the feedback FF-IC (with finite logarithmic Jensen's gap $c_{JG}$)
with feedback and delayed CSIT can be proved to be same as the capacity
region of a feedback IC (without fading) with equivalent channel strengths
$SNR_{i}:=\expect{\abs{g_{ii}}^{2}}$ for $i=1,2$, and $INR_{i}:=\expect{\abs{g_{ij}}^{2}}$
for $i\neq j$. \label{cor:static=00003Dfading_fb}
\end{cor}
\begin{IEEEproof}
This is again an application of the logarithmic Jensen's gap result.
The proof is given in \ifarxiv  Appendix \ref{app:proof_cor_fading_is_static_fB}\else  \cite[Appendix F]{Joyson_jensens_gap_arxiv}\fi.
\end{IEEEproof}

\subsection{Proof of Theorem~\ref{th:splitting-fb}\label{subsec:fb}}

Note that since the receivers know their respective incoming channel
states, we can view the effective channel output at Rx$i$ as the
pair $\left(Y_{i},\underline{g}_{i}\right)$. Then the block Markov
scheme of \cite[Lemma 1]{suh_tse_fb_gaussian} implies that the rate
pairs $(R_{1},R_{2})$ satisfying \begin{subequations}\label{eq:ach_fb}
\begin{align}
R_{1} & \leq I\brac{U,U_{2},X_{1};Y_{1},\underline{g_{1}}}\label{eq:ach_fb1}\\
R_{1} & \leq I\brac{U_{1};Y_{2},\underline{g_{2}}|U,X_{2}}+I\brac{X_{1};Y_{1},\underline{g_{1}}|U_{1},U_{2},U}\label{eq:ach_fb2}\\
R_{2} & \leq I\brac{U,U_{1},X_{2};Y_{2},\underline{g_{2}}}\label{eq:ach_fb3}\\
R_{2} & \leq I\brac{U_{2};Y_{1},\underline{g_{1}}|U,X_{1}}+I\brac{X_{2};Y_{2},\underline{g_{2}}|U_{1},U_{2},U}\label{eq:ach_fb4}\\
R_{1}+R_{2} & \leq I\brac{X_{1};Y_{1},\underline{g_{1}}|U_{1},U_{2},U}+I\brac{U,U_{1},X_{2};Y_{2},\underline{g_{2}}}\label{eq:ach_fb5}\\
R_{1}+R_{2} & \leq I\brac{X_{2};Y_{2},\underline{g_{2}}|U_{1},U_{2},U}+I\brac{U,U_{2},X_{1};Y_{1},\underline{g_{1}}}\label{eq:ach_fb6}
\end{align}
\end{subequations}for all $p\brac up\brac{u_{1}|u}p\brac{u_{2}|u}p\brac{x_{1}|u_{1},u}p\brac{x_{2}|u_{2},u}$
are achievable. We choose the input distribution according to
\begin{equation}
U\sim\mathcal{CN}\brac{0,\abs{\rho}^{2}},U_{k}\sim\mathcal{CN}\brac{0,\lambda_{ck}},X_{pk}\sim\mathcal{CN}\brac{0,\lambda_{pk}}
\end{equation}
\begin{equation}
X_{1}=e^{i\theta}U+U_{1}+X_{p1},\quad X_{2}=U+U_{2}+X_{p2}
\end{equation}
($U,U_{k},X_{pk}$ being mutually independent) with $0\leq\abs{\rho}^{2}\leq1$,
$0\leq\theta<2\pi$, $\lambda_{ck}+\lambda_{pk}=1-\abs{\rho}^{2}$
and $\lambda_{pk}=\min\brac{\frac{1}{INR_{k}},1-\abs{\rho}^{2}}$.

With this choice of $\lambda_{pk}$ we perform the rate-splitting
according to the average $inr$ in place of rate-splitting based on
the constant $inr$. Note that we have introduced an extra rotation
$\theta$ for the first transmitter, which will become helpful in
proving the capacity gap by allowing us to choose a point in inner
bound for every point in outer bound (see proof of claim \ref{claim:gap_fb}).
On evaluating the terms in \eqref{eq:ach_fb} for this choice of input
distribution, we get the inner bound described by \eqref{eq:inner_fb};
the calculations are deferred to \ifarxiv  Appendix \ref{app:achievability_fb}\else  \cite[Appendix G]{Joyson_jensens_gap_arxiv}\fi.

An outer bound for the feedback case is given by \eqref{eq:outer_fb}
with $0\le\abs{\rho}\leq1$ ($\rho$ being a complex number): \begin{subequations}\label{eq:outer_fb}
\begin{align}
R_{1} & \leq\expect{\lgbrac{\abs{g_{11}}^{2}+\abs{g_{21}}^{2}+2\text{Re}\brac{\rho g_{11}g_{21}^{*}}+1}}\label{eq:outer_fb1}\\
R_{1} & \leq\expect{\lgbrac{1+\brac{1-\abs{\rho}^{2}}\abs{g_{12}}^{2}}}+\expect{\lgbrac{1+\frac{\brac{1-\abs{\rho}^{2}}\abs{g_{11}}^{2}}{1+\brac{1-\abs{\rho}^{2}}\abs{g_{12}}^{2}}}}\label{eq:outer_fb2}\\
R_{2} & \leq\expect{\lgbrac{\abs{g_{22}}^{2}+\abs{g_{12}}^{2}+2\text{Re}\brac{\rho g_{22}^{*}g_{12}}+1}}\label{eq:outer_fb3}\\
R_{2} & \leq\expect{\lgbrac{1+\brac{1-\abs{\rho}^{2}}\abs{g_{21}}^{2}}}+\expect{\lgbrac{1+\frac{\brac{1-\abs{\rho}^{2}}\abs{g_{22}}^{2}}{1+\brac{1-\abs{\rho}^{2}}\abs{g_{21}}^{2}}}}\label{eq:outer_fb4}\\
R_{1}+R_{2} & \leq\expect{\lgbrac{\abs{g_{22}}^{2}+\abs{g_{12}}^{2}+2\text{Re}\brac{\rho g_{22}^{*}g_{12}}+1}}\nonumber \\
 & \qquad+\expect{\lgbrac{1+\frac{\brac{1-\abs{\rho}^{2}}\abs{g_{11}}^{2}}{1+\brac{1-\abs{\rho}^{2}}\abs{g_{12}}^{2}}}}\label{eq:outer_fb5}\\
R_{1}+R_{2} & \leq\expect{\lgbrac{\abs{g_{11}}^{2}+\abs{g_{21}}^{2}+2\text{Re}\brac{\rho g_{11}g_{21}^{*}}+1}}\nonumber \\
 & \qquad+\expect{\lgbrac{1+\frac{\brac{1-\abs{\rho}^{2}}\abs{g_{22}}^{2}}{1+\brac{1-\abs{\rho}^{2}}\abs{g_{21}}^{2}}}}.\label{eq:outer_fb6}
\end{align}
\end{subequations}

The outer bounds can be easily derived following the proof techniques
from \cite[Theorem 3]{suh_tse_fb_gaussian} using $\expect{X_{1}X_{2}^{*}}=\rho$,
treating $\left(Y_{i},\underline{g}_{i}\right)$ as output, and using
the i.i.d property of the channels. The calculations are deferred
to \ifarxiv  Appendix \ref{app:outer_bounds_fb}\else  \cite[Appendix H]{Joyson_jensens_gap_arxiv}\fi.
\begin{claim}
The gap between the inner bound \eqref{eq:inner_fb} and the outer
bound \eqref{eq:outer_fb} for the feedback case is at most $c_{JG}+2$
bits per channel use.\label{claim:gap_fb}
\end{claim}
\begin{IEEEproof}
Denote the gap between the first outer bound $(\ref{eq:outer_fb1})$
and inner bound $(\ref{eq:inner_fb1})$ by $\delta_{1}$, for the
second pair denote the gap by $\delta_{2}$, and so on. For comparing
the gap between regions we choose the inner bound point with same
$\abs{\rho}$ as in any given outer bound point. The rotation $\theta$
for the first transmitter also becomes important in proving a constant
gap capacity result. We choose $\theta$ in the inner bound to match
$\arg\brac{\rho}$ in the outer bound. With this choice we get
\begin{align}
\delta_{1} & =\expect{\lgbrac{\abs{g_{11}}^{2}+\abs{g_{21}}^{2}+2\abs{\rho}\text{Re}\brac{e^{i\theta}g_{11}g_{21}^{*}}+1}}\nonumber \\
 & \quad-\expect{\lgbrac{\abs{g_{11}}^{2}+\abs{g_{21}}^{2}+2\abs{\rho}^{2}\text{Re}\brac{e^{i\theta}g_{11}g_{21}^{*}}+1}}+1\\
 & =\expect{\lgbrac{\frac{1+\frac{1}{\abs{g_{11}}^{2}+\abs{g_{21}}^{2}}+\abs{\rho}\brac{\frac{2\text{Re}\brac{e^{i\theta}g_{11}g_{21}^{*}}}{\abs{g_{11}}^{2}+\abs{g_{21}}^{2}}}}{1+\frac{1}{\abs{g_{11}}^{2}+\abs{g_{21}}^{2}}+\abs{\rho}^{2}\brac{\frac{2\text{Re}\brac{e^{i\theta}g_{11}g_{21}^{*}}}{\abs{g_{11}}^{2}+\abs{g_{21}}^{2}}}}}}+1.
\end{align}
We have $\abs{\frac{2\text{Re}\brac{e^{i\theta}g_{11}g_{21}^{*}}}{\abs{g_{11}}^{2}+\abs{g_{21}}^{2}}}=\frac{\abs{e^{-i\theta}g_{11}^{*}g_{21}+e^{i\theta}g_{11}g_{21}^{*}}}{\abs{g_{11}}^{2}+\abs{g_{21}}^{2}}\leq1,$
hence we call $\frac{e^{-i\theta}g_{11}^{*}g_{21}+e^{i\theta}g_{11}g_{21}^{*}}{\abs{g_{11}}^{2}+\abs{g_{21}}^{2}}=\sin\varphi$
and let $\abs{g_{11}}^{2}+\abs{g_{21}}^{2}=r^{2}.$ Therefore,
\begin{align}
\delta_{1} & =\expect{\lgbrac{\frac{1+\frac{1}{r^{2}}+\abs{\rho}\sin\varphi}{1+\frac{1}{r^{2}}+\abs{\rho}^{2}\sin\varphi}}}+1.
\end{align}
If $\sin\phi<0$, then $\frac{1+\frac{1}{r^{2}}+\abs{\rho}\sin\varphi}{1+\frac{1}{r^{2}}+\abs{\rho}^{2}\sin\varphi}\leq1.$
Otherwise, if $\sin\phi>0$, then $\frac{1+\frac{1}{r^{2}}+\abs{\rho}\sin\varphi}{1+\frac{1}{r^{2}}+\abs{\rho}^{2}\sin\varphi}=1+\frac{\brac{\abs{\rho}-\abs{\rho}^{2}}\sin\varphi}{1+\frac{1}{r^{2}}+\abs{\rho}^{2}\sin\varphi}\leq2$
since $0\leq\brac{\abs{\rho}-\abs{\rho}^{2}}\sin\phi\leq1$ and $1+\frac{1}{r^{2}}+\abs{\rho}^{2}\sin\phi>1$.
Hence, we always get
\begin{equation}
\delta_{1}\leq2.
\end{equation}

Now we consider the gap $\delta_{2}$ between the second inequality
$(\ref{eq:outer_fb2})$ of the outer bound and the second inequality
$(\ref{eq:inner_fb2})$ of the inner bound.
\begin{align}
\delta_{2} & =\expect{\lgbrac{1+\frac{\brac{1-\abs{\rho}^{2}}\abs{g_{11}}^{2}}{1+\brac{1-\abs{\rho}^{2}}\abs{g_{12}}^{2}}}}-\expect{\lgbrac{1+\lambda_{p1}\abs{g_{11}}^{2}+\lambda_{p2}\abs{g_{21}}^{2}}}+2\\
 & \overset{(a)}{\leq}\expect{\lgbrac{1+\brac{1-\abs{\rho}^{2}}INR_{1}+\brac{1-\abs{\rho}^{2}}\abs{g_{11}}^{2}}}-\lgbrac{1+\brac{1-\abs{\rho}^{2}}INR_{1}}+c_{JG}\nonumber \\
 & \qquad-\expect{\lgbrac{1+\lambda_{p1}\abs{g_{11}}^{2}+\lambda_{p2}\abs{g_{21}}^{2}}}+2\\
 & \leq\expect{\lgbrac{1+\frac{\brac{1-\abs{\rho}^{2}}\abs{g_{11}}^{2}}{1+\brac{1-\abs{\rho}^{2}}INR_{1}}}}-\expect{\lgbrac{1+\lambda_{p1}\abs{g_{11}}^{2}}}+2+c_{JG}\\
 & \overset{(b)}{\leq}2+c_{JG},
\end{align}
where $(a)$ follows by using logarithmic Jensen's gap property on
$\abs{g_{12}}^{2}$ and Jensen's inequality. The step $(b)$ follows
because
\begin{equation}
\frac{\brac{1-\abs{\rho}^{2}}}{1+\brac{1-\abs{\rho}^{2}}INR_{1}}=\frac{1}{\frac{1}{1-\abs{\rho}^{2}}+INR_{1}}\leq\min\brac{\frac{1}{INR_{1}},1-\abs{\rho}^{2}}=\lambda_{p1}.
\end{equation}
Similarly, by inspection of the other bounding inequalities we can
gather the inequalities on the $\delta$'s as:
\begin{equation}
\delta_{1},\delta_{3}\leq2;\quad\delta_{2},\delta_{4}\leq c_{JG}+2;\quad\delta_{5},\delta_{6}\leq c_{JG}+3.
\end{equation}
We note that $\brac{\delta_{1},\delta_{2}}$ is associated with bounding
$R_{1}$, $\brac{\delta_{3},\delta_{4}}$ with $R_{2}$, $\brac{\delta_{5},\delta_{6}}$
with $R_{1}+R_{2}$. Hence it follows that the capacity gap is at
most $\max\brac{\delta_{1},\delta_{2},\delta_{3},\delta_{4},\frac{\delta_{5}}{2},\frac{\delta_{6}}{2}}\leq c_{JG}+2$
bits per channel use.
\end{IEEEproof}

\section{Approximate capacity of feedback FF-IC using point-to-point codes\label{sec:FF_IC_p2p_codes}}

As the third illustration for the usefulness of logarithmic Jensen's
gap, we propose a strategy that does not make use of rate-splitting,
superposition coding or joint decoding for the feedback case, which
achieves the entire capacity region for 2-user symmetric FF-ICs to
within a constant gap. This constant gap is dictated by the logarithmic
Jensen's gap for the fading model. Our scheme only uses point-to-point
codes, and a feedback scheme based on amplify-and-forward relaying,
similar to the one proposed in \cite{suh_tse_fb_gaussian}.

The main idea behind the scheme is to have one of the transmitters
initially send a very densely modulated block of data, and then refine
this information using feedback and amplify-and-forward relaying for
the following blocks, in a fashion similar to the Schalkwijk-Kailath
scheme \cite{sk_66}, while treating the interference as noise. Such
refinement effectively induces a 2-tap point-to-point inter-symbol-interference
(ISI) channel at the unintended receiver, and a point-to-point feedback
channel for the intended receiver. As a result, both receivers can
decode their intended information using only point-to-point codes.

Consider the symmetric FF-IC, where the channel statistics are symmetric
and independent, \emph{i.e.}, $g_{11}(l)\sim g_{22}(l)\sim g_{d}$
and $g_{12}(l)\sim g_{21}(l)\sim g_{c}$ and all the random variables
independent of each other. We consider $n$ transmission phases, each
phase having a block length of $N$. For Tx$1$, generate $2^{nNR_{1}}$
codewords \allowbreak $\brac{X_{1}^{(1)N},\ldots,X_{1}^{(n)N}}$
i.i.d according to $\mathcal{CN}\brac{0,1}$ and encode its message
$W_{1}\in\cbrac{1,\ldots,2^{nNR_{1}}}$ onto \allowbreak $\brac{X_{1}^{(1)N},\ldots,X_{1}^{(n)N}}.$
For Tx$2$, generate $2^{nNR_{2}}$ codewords $X_{2}^{(1)N}=X_{2}^{N}$
i.i.d according to $\mathcal{CN}\brac{0,1}$ and encode its message
$W_{2}\in\cbrac{1,\ldots,2^{nNR_{2}}}$ onto $X_{2}^{(1)N}=X_{2}^{N}$.
Note that for Tx$2$ the message is encoded into $N$ length sequence
to be transmitted at first phase, whereas for Tx$1$ the message is
encoded into $nN$ length sequence to be transmitted through $n$
phases.

Tx$1$  sends $X_{1}^{(i)N}$ in phase $i$. Tx$2$  sends $X_{2}^{(1)N}=X_{2}^{N}$
in phase 1. At the beginning of phase $i>1$, Tx$2$  receives
\begin{equation}
Y_{2}^{(i-1)N}=g_{22}^{(i-1)N}X_{2}^{(i-1)N}+g_{12}^{(i-1)N}X_{1}^{(i-1)N}+Z_{2}^{(i-1)N}
\end{equation}
from feedback. It can remove $g_{22}^{(i-1)N}X_{2}^{(i-1)N}$ from
$Y_{2}^{(i-1)N}$ to obtain $g_{12}^{(i-1)N}X_{1}^{(i-1)N}+Z_{2}^{(i-1)N}$.
Tx$2$  then transmits the resulting interference-plus-noise after
power scaling as $X_{2}^{(i)N}$, i.e.
\begin{equation}
X_{2}^{(i)N}=\frac{g_{12}^{(i-1)N}X_{1}^{(i-1)N}+Z_{2}^{(i-1)N}}{\sqrt{1+INR}}.
\end{equation}
Thus, in phase $i>1$, Rx$2$  receives
\begin{align}
Y_{2}^{\brac iN} & =g_{22}^{(i)N}X_{2}^{(i)N}+g_{12}^{(i)N}X_{1}^{(i)N}+Z_{2}^{(i)N}\\
 & =g_{22}^{(i)N}\brac{\frac{g_{12}^{(i-1)N}X_{1}^{(i-1)N}+Z_{2}^{(i-1)N}}{\sqrt{1+INR}}}+g_{12}^{(i)N}X_{1}^{(i)N}+Z_{2}^{(i)N}
\end{align}
and feeds it back to Tx$2$  for phase $i+1$. The transmission scheme
is summarized in Table \ref{tab:n-Phase-scheme}. Note that for phase
$i=1$ Tx$1$ receives
\begin{alignat}{1}
Y_{1}^{\brac 1N} & =g_{11}^{(1)N}X_{1}^{(1)N}+g_{21}^{(1)N}X_{2}^{(1)N}+Z_{1}^{(1)N}
\end{alignat}
and for phase $i>1$ Tx$1$  observes a block ISI channel since it
receives
\begin{alignat}{1}
Y_{1}^{\brac iN} & =g_{11}^{(i)N}X_{1}^{(i)N}+g_{21}^{(i)N}\brac{\frac{g_{12}^{(i-1)N}X_{1}^{(i-1)N}+Z_{2}^{(i-1)N}}{\sqrt{1+INR}}}+Z_{1}^{(i)N}\label{eq:isi_1}\\
 & =g_{11}^{(i)N}X_{1}^{(i)N}+\brac{\frac{g_{21}^{(i)N}g_{12}^{(i-1)N}}{\sqrt{1+INR}}}X_{1}^{(i-1)N}+\tilde{Z}_{1}^{(i)N},\label{eq:isi_2}
\end{alignat}
where $\tilde{Z}_{1}^{(i)N}=Z_{1}^{(i)N}+\frac{g_{21}^{(i)N}Z_{2}^{(i-1)N}}{\sqrt{1+INR}}$.

\begin{table}[htbp]
\centering{}\caption{\label{tab:n-Phase-scheme} Transmitted symbols in $n$-phase scheme
for symmetric FF-IC with feedback}
\begin{tabular}{|c|c|c|c|c|c|}
\hline
User  & Phase $1$  & Phase $2$  & .  & .  & Phase $n$\tabularnewline
\hline
\hline
1  & $X_{1}^{(1)N}$  & $X_{1}^{(2)N}$  & .  & .  & $X_{1}^{(n)N}$\tabularnewline
\hline
2  & $X_{2}^{(1)N}$  & $\frac{g_{12}^{(1)N}X_{1}^{(1)N}+Z_{2}^{(1)N}}{\sqrt{1+INR}}$  & .  & .  & $\frac{g_{12}^{(n-1)N}X_{1}^{(n-1)N}+Z_{2}^{(n-1)N}}{\sqrt{1+INR}}$\tabularnewline
\hline
\end{tabular}
\end{table}

At the end of $n$ blocks, Rx$1$ collects ${\bf Y_{1}}^{N}=\brac{Y_{1}^{(1)N},\ldots,Y_{1}^{(n)N}}$
and decodes $W_{1}$ such that $\brac{{\bf X_{1}}^{N}\brac{W_{1}},{\bf Y_{1}}^{N}}$
is jointly typical (where ${\bf X_{1}}^{N}=\brac{X_{1}^{(1)N},\ldots,X_{1}^{(n)N}}$)
treating $X_{2}^{(1)N}=X_{2}^{N}$ as noise. At Rx$2$, channel outputs
over $n$ phases can be combined with an appropriate scaling so that
the interference-plus-noise at phases $\{1,\dots,n-1\}$ are successively
canceled, \emph{i.e.}, an effective point-to-point channel can be
generated through $\tilde{Y}_{2}^{N}=Y_{2}^{\brac nN}+\sum_{i=1}^{n-1}\brac{\prod_{j=i+1}^{n}\frac{-g_{22}^{(j)N}}{\sqrt{1+INR}}}Y_{2}^{\brac iN}$
(see the analysis in the subsection \ref{subsec:p2p_for_sym_fad_ics}
for details). Note that this can be viewed as a block version of the
Schalkwijk-Kailath scheme \cite{sk_66}. Given the effective channel
$\tilde{Y}_{2}^{N}$, the receiver can simply use point-to-point typicality
decoding to recover $W_{2}$, treating the interference in phase $n$
as noise.
\begin{thm}
For a symmetric FF-IC with a finite logarithmic Jensen's gap $c_{JG}$
, the rate pair
\begin{align*}
\left(R_{1},R_{2}\right)=\left(\lgbrac{1+SNR+INR}-3c_{JG}-2,\expect{\log^{+}\left[\frac{\abs{g_{d}}^{2}}{1+INR}\right]}\right)
\end{align*}
is achievable by the scheme. The scheme together with switching the
roles of users and time-sharing, achieves the capacity region of symmetric
feedback IC within $3c_{JG}+2$ bits per channel use.
\end{thm}
\begin{IEEEproof}
The proof follows from the analysis in the following subsection.
\end{IEEEproof}

\subsection{Analysis of Point-to-Point Codes for Symmetric FF-ICs\label{subsec:p2p_for_sym_fad_ics}}

We now provide the analysis for the scheme, going through the decoding
at the two receivers and then looking at the capacity gap for the
achievable region.

\subsubsection{Decoding at Rx$1$ }

At the end of $n$ blocks Rx$1$  collects ${\bf Y_{1}}^{N}=\brac{Y_{1}^{(1)N},\ldots Y_{1}^{(n)N}}$
and decodes $W_{1}$ such that $\brac{{\bf X_{1}}^{N}\brac{W_{1}},{\bf Y_{1}}^{N}}$
is jointly typical, where ${\bf X_{1}}^{N}=\brac{X_{1}^{(1)N},\ldots X_{1}^{(n)N}}$.
The joint typicality is considered according the product distribution
$p^{N}\brac{{\bf X_{1}},{\bf Y_{1}}}$, where \allowbreak
\begin{equation}
p\brac{{\bf X_{1}},{\bf Y_{1}}}=p\brac{\brac{X_{1}^{(1)},\ldots X_{1}^{(n)}},\brac{Y_{1}^{(1)},\ldots Y_{1}^{(n)}}}
\end{equation}
 is a joint Gaussian distribution, dictated by the following equations
that arise from our $n$-phase scheme:
\begin{alignat}{1}
Y_{1}^{\brac 1} & =g_{11}^{(1)}X_{1}^{(1)}+g_{21}^{(1)}X_{2}^{(1)}+Z_{1}^{(1)}.
\end{alignat}
And for $i=2,3,\ldots,n$:
\begin{alignat}{1}
Y_{1}^{\brac i} & =g_{11}^{(i)}X_{1}^{(i)}+g_{21}^{(i)}\brac{\frac{g_{12}^{(i-1)}X_{1}^{(i-1)}+Z_{2}^{(i-1)}}{\sqrt{1+INR}}}+Z_{1}^{(i)}
\end{alignat}
with $X_{1}^{(i)},X_{2}^{(1)},Z_{1}^{(i)},Z_{2}^{(i-1)}$ being i.i.d
$\mathcal{CN}\brac{0,1}$. Essentially $X_{2}^{(1)},Z_{1}^{(i)}$
are both Gaussian noise for Rx1.

Using standard techniques it follows that for the $n$-phase scheme,
as $N\rightarrow\infty$ user 1 can achieve the rate $\frac{1}{n}\expect{\lgbrac{\frac{\abs{K_{{\bf Y_{1}}}(n)}}{\abs{K_{{\bf Y_{1}|X_{1}}}(n)}}}}$,
where $\abs{K_{{\bf Y_{1}}}(n)}$ denotes the determinant of covariance
matrix for the $n$-phase scheme defined in the following pattern:{\small{}
\[
K_{{\bf Y_{1}}}(1)=\left[1+\abs{g_{11}\brac 1}^{2}+\abs{g_{21}\brac 1}^{2}\right],
\]
\[
K_{{\bf Y_{1}}}(2)=\left[\begin{array}{cc}
\abs{g_{11}\brac 2}^{2}+\frac{\abs{g_{21}\brac 2}^{2}\brac{\abs{g_{12}\brac 1}^{2}+1}}{1+INR}+1 & \frac{g_{11}^{*}\brac 1g_{21}\brac 2g_{12}\brac 1}{\sqrt{1+INR}}\\
\frac{g_{11}\brac 1g_{21}^{*}\brac 2g_{12}^{*}\brac 1}{\sqrt{1+INR}} & \abs{g_{11}\brac 1}^{2}+\abs{g_{21}\brac 1}^{2}+1
\end{array}\right],
\]
}{\small \par}

{\small{}}%
{\small{}
\[
K_{{\bf Y_{1}}}(l)=\left[\begin{array}{cc}
\abs{g_{11}\brac l}^{2}+\frac{\abs{g_{21}\brac l}^{2}\brac{\abs{g_{12}\brac{l-1}}^{2}+1}}{1+INR}+1 & \sbrac{\frac{g_{11}^{*}\brac{l-1}g_{21}\brac lg_{12}\brac{l-1}}{\sqrt{1+INR}},0_{l-2}}\\
\sbrac{\frac{g_{11}^{*}\brac{l-1}g_{21}\brac lg_{12}\brac{l-1}}{\sqrt{1+INR}},0_{l-2}}^{\dagger} & K_{{\bf Y_{1}}}(l-1)
\end{array}\right],
\]
}where $0_{l-2}$ is $\brac{l-2}$ length zero vector, $\dagger$
indicates Hermitian conjugate, $g_{11}\brac i\sim g_{d}$ i.i.d and
$g_{12}\brac i,g_{21}\brac i\sim g_{c}$ i.i.d. Letting $n\rightarrow\infty$,
Rx$1$  can achieve the rate $R_{1}=\underset{n\rightarrow\infty}{\text{lim}}\frac{1}{n}\expect{\lgbrac{\frac{\abs{K_{{\bf Y_{1}}}(n)}}{\abs{K_{{\bf Y_{1}|X_{1}}}(n)}}}}$.
We need to evaluate $\underset{n\rightarrow\infty}{\text{lim}}\frac{1}{n}\expect{\lgbrac{\frac{\abs{K_{{\bf Y_{1}}}(n)}}{\abs{K_{{\bf Y_{1}|X_{1}}}(n)}}}}$.
The following lemma gives a lower bound on $\frac{1}{n}\expect{\lgbrac{\abs{K_{{\bf Y_{1}}}(n)}}}$.
\begin{lem}
\begin{align*}
\frac{1}{n}\expect{\lgbrac{\abs{K_{{\bf Y_{1}}}(n)}}} & \geq\frac{1}{n}\lgbrac{\abs{\hat{K}_{{\bf Y_{1}}}(n)}}-3c_{JG},
\end{align*}
\label{lem:fading_matrix}where $\hat{K}_{{\bf Y_{1}}}(n)$ is obtained
from $K_{{\bf Y_{1}}}(n)$ by replacing $g_{12}\brac i$'s,$g_{21}\brac i$'s
with $\sqrt{INR}$ and $g_{11}\brac i$'s with $\sqrt{SNR}$.
\end{lem}
\begin{IEEEproof}
The proof involves expanding the matrix determinant and repeated application
of the logarithmic Jensen's gap property. The details are given in
\ifarxiv  Appendix \ref{app:Fadingmatrix}\else  \cite[Appendix I]{Joyson_jensens_gap_arxiv}\fi.
\end{IEEEproof}
Subsequently, we use the following lemma in bounding $\underset{n\rightarrow\infty}{\text{lim}}\frac{1}{n}\lgbrac{\abs{\hat{K}_{{\bf Y_{1}}}(n)}}$.
\begin{lem}
\label{lem:detlemma}If $A_{1}=\left[\abs a\right],A_{2}=\left[\begin{array}{cc}
\abs a & b\\
b^{*} & \abs a
\end{array}\right],A_{3}=\left[\begin{array}{ccc}
\abs a & b & 0\\
b^{*} & \abs a & b\\
0 & b^{*} & \abs a
\end{array}\right],$ %
{} etc. with $\abs a^{2}>4\abs b^{2}$, then
\[
\underset{n\rightarrow\infty}{\liminf}\frac{1}{n}\lgbrac{\abs{A_{n}}}\geq\lgbrac{\abs a}-1.
\]
\end{lem}
\begin{IEEEproof}
The proof is given in \ifarxiv  Appendix \ref{app:proof_matrix_asym}\else  \cite[Appendix J]{Joyson_jensens_gap_arxiv}\fi.
\end{IEEEproof}
For the $n$-phase scheme, the $\abs{\hat{K}_{{\bf Y_{1}}}(n)}$ matrix
has the form $A_{n}$, as defined in Lemma \ref{lem:detlemma} after
identifying $\abs a=1+INR+SNR$ and $b=\frac{\sqrt{SNR}INR}{\sqrt{1+INR}}$.
Note that with this choice $\abs a^{2}>4\abs b^{2}$ holds due to
AM-GM (Arithmetic Mean $\geq$ Geometric Mean)  inequality. Hence,
we have
\begin{equation}
\underset{n\rightarrow\infty}{\liminf}\frac{1}{n}\lgbrac{\abs{\hat{K}_{{\bf Y_{1}}}(n)}}\geq\lgbrac{1+INR+SNR}-1\label{eq:p2p_1}
\end{equation}
using Lemma \ref{lem:detlemma}. Also, $K_{{\bf Y_{1}|X_{1}}}(n)$
is a diagonal matrix of the form

\begin{equation}
K_{{\bf Y_{1}|X_{1}}}(n)=\text{diag}\brac{\frac{\abs{g_{21}\brac n}^{2}}{1+INR}+1,\frac{\abs{g_{21}\brac{n-1}}^{2}}{1+INR}+1,\ldots,\frac{\abs{g_{21}\brac 2}^{2}}{1+INR}+1,\abs{g_{21}\brac 1}^{2}+1}.
\end{equation}
Hence, using Jensen's inequality
\begin{align}
\underset{n\rightarrow\infty}{\limsup}\frac{1}{n}\expect{\lgbrac{\abs{K_{{\bf Y_{1}|X_{1}}}(n)}}} & \leq\underset{n\rightarrow\infty}{\limsup}\frac{1}{n}\lgbrac{\brac{\frac{INR}{1+INR}+1}^{n-1}\brac{1+INR}}\\
 & =\lgbrac{\frac{INR}{1+INR}+1}\\
 & \leq1.\label{eq:p2p_2}
\end{align}
Hence, by combining Lemma \ref{lem:fading_matrix}, Equation \eqref{eq:p2p_1}
and Equation \eqref{eq:p2p_2}, we get
\begin{alignat}{1}
R_{1} & \leq\lgbrac{1+INR+SNR}-3c_{JG}-2
\end{alignat}
is achievable.

\subsubsection{Decoding at Rx$2$ \label{subsec:decoding_rx2_sk}}

For user 2 we can use a block variant of Schalkwijk-Kailath scheme
\cite{sk_66} to achieve $R_{2}=\expect{\log^{+}\left(\frac{\abs{g_{d}}^{2}}{1+INR}\right)}$.
The key idea is that the interference-plus-noise sent in subsequent
slots can indeed refine the symbols of the previous slot. The chain
of refinement over $n$ phases compensate for the fact that the information
symbols are sent only in the first phase. We have
\begin{equation}
Y_{2}^{\brac 1N}=g_{22}^{(1)N}X_{2}^{N}+g_{12}^{(1)N}X_{1}^{(1)N}+Z_{2}^{(1)N}
\end{equation}
and
\begin{equation}
Y_{2}^{\brac iN}=g_{22}^{(i)N}\brac{\frac{g_{12}^{(i-1)N}X_{1}^{(i-1)N}+Z_{2}^{(i-1)N}}{\sqrt{1+INR}}}+g_{12}^{(i)N}X_{1}^{(i)N}+Z_{2}^{(i)N}
\end{equation}
for $i>1$. Now let $\tilde{Y}_{2}^{N}=Y_{2}^{\brac nN}+\sum_{i=1}^{n-1}\brac{\prod_{j=i+1}^{n}\frac{-g_{22}^{(j)N}}{\sqrt{1+INR}}}Y_{2}^{\brac iN}$.
We have
\begin{alignat}{1}
\tilde{Y}_{2}^{N} & =Y_{2}^{\brac nN}+\sum_{i=1}^{n-1}\brac{\prod_{j=i+1}^{n}\frac{-g_{22}^{(j)N}}{\sqrt{1+INR}}}Y_{2}^{\brac iN}\\
 & =g_{22}^{(n)N}\brac{\frac{g_{12}^{(n-1)N}X_{1}^{(n-1)N}+Z_{2}^{(n-1)N}}{\sqrt{1+INR}}}+g_{12}^{(n)N}X_{1}^{(n)N}+Z_{2}^{(n)N}\nonumber \\
 & \qquad+\brac{\frac{-g_{22}^{(n)N}}{\sqrt{1+INR}}}\brac{g_{22}^{(n-1)N}\brac{\frac{g_{12}^{(n-2)N}X_{1}^{(n-2)N}+Z_{2}^{(n-2)N}}{\sqrt{1+INR}}}+g_{12}^{(n-1)N}X_{1}^{(n-1)N}+Z_{2}^{(n-1)N}}\nonumber \\
 & \qquad+\brac{\frac{g_{22}^{(n)N}g_{22}^{(n-1)N}}{1+INR}}\brac{g_{22}^{(n-2)N}\brac{\frac{g_{12}^{(n-3)N}X_{1}^{(n-3)N}+Z_{2}^{(n-3)N}}{\sqrt{1+INR}}}+g_{12}^{(n-2)N}X_{1}^{(n-2)N}+Z_{2}^{(n-2)N}}\nonumber \\
 & \qquad+\cdots\nonumber \\
 & \qquad+\brac{\prod_{j=2}^{n}\frac{-g_{22}^{(j)N}}{\sqrt{1+INR}}}\brac{g_{22}^{(1)N}X_{2}^{N}+g_{21}^{(1)N}X_{2}^{(1)N}+Z_{1}^{(1)N}}\\
 & =g_{22}^{(1)N}\brac{\prod_{j=2}^{n}\frac{-g_{22}^{(j)N}}{\sqrt{1+INR}}}X_{2}^{N}+g_{12}^{(n)N}X_{1}^{(n)N}+Z_{2}^{(n)N}.
\end{alignat}
due to cross-cancellation. Now Rx$2$  decodes for its message from
$\tilde{Y}_{2}^{N}$. Hence, Rx$2$  can achieve the rate
\begin{alignat}{1}
R_{2} & \leq\underset{n\rightarrow\infty}{\liminf}\frac{1}{n}\expect{\lgbrac{1+\brac{\prod_{j=2}^{n}\frac{\abs{g_{22}\brac j}^{2}}{1+INR}}\frac{\abs{g_{22}\brac 1}^{2}}{1+\abs{g_{12}\brac n}^{2}}}},
\end{alignat}
where $g_{22}\brac 1,\ldots,g_{22}\brac n\sim g_{d}$ being i.i.d
and $g_{12}\brac n\sim g_{c}.$ Hence, it follows that
\begin{equation}
R_{2}\leq\expect{\log^{+}\left(\frac{\abs{g_{d}}^{2}}{1+INR}\right)}
\end{equation}
is achievable.

\subsubsection{Capacity gap}

\begin{figure}[!th]
\centering{}\includegraphics[scale=0.6]{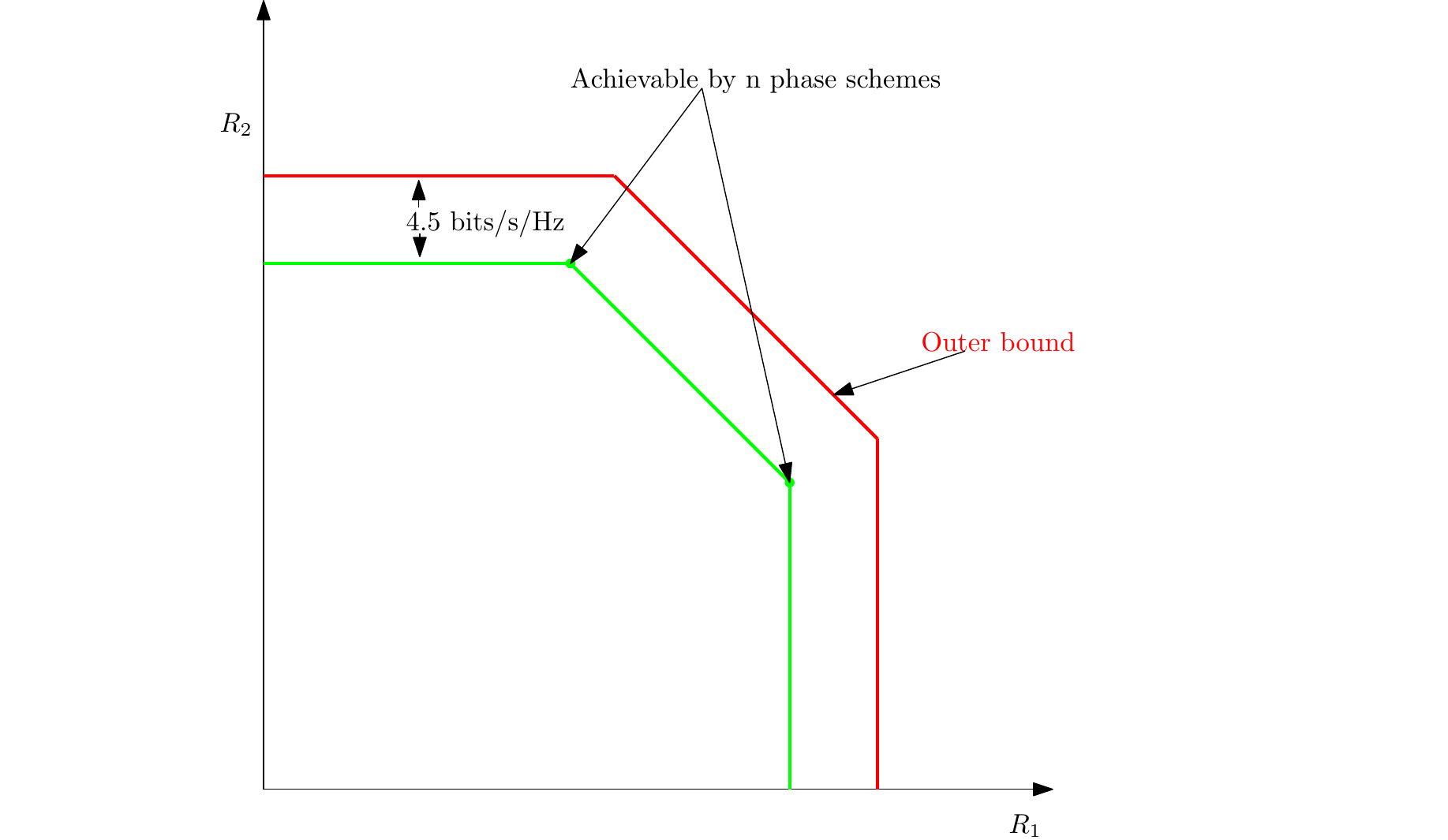}\caption{Illustration of bounds for capacity region for symmetric FF-IC. The
corner points of the outer bound can be approximately achieved by
our $n$-phase schemes. The gap is approximately $4.5$ bits per channel
use for the Rayleigh fading case. \label{fig:fb_IC_capacity}}
\end{figure}

We can obtain the following outer bounds from Theorem \ref{thm:apprx_capacity_fb}
for the special case of symmetric fading statistics.
\begin{align}
R_{1},R_{2} & \leq\expect{\lgbrac{\abs{g_{d}}^{2}+\abs{g_{c}}^{2}+1}}\label{eq:outer_bound_sym_fb1}\\
R_{1}+R_{2} & \leq\expect{\lgbrac{1+\frac{\abs{g_{d}}^{2}}{1+\abs{g_{c}}^{2}}}}+\expect{\lgbrac{\abs{g_{d}}^{2}+\abs{g_{c}}^{2}+2\abs{g_{d}}\abs{g_{c}}+1}},\label{eq:outer_bound_sym_fb2}
\end{align}
where Equation \eqref{eq:outer_bound_sym_fb1} is obtained from Equation
\eqref{eq:outer_fb2} and Equation \eqref{eq:outer_fb4} by setting
$\rho=0$ (note that $\rho=0$ yields the loosest version of outer
bounds in Equation \eqref{eq:outer_fb2} and Equation \eqref{eq:outer_fb4}).
Similarly, Equation \eqref{eq:outer_bound_sym_fb2} is a looser version
of outer bound Equation \eqref{eq:outer_fb5} independent of $\rho$.
The outer bounds reduce to a pentagonal region with two non-trivial
corner points (see Figure \ref{fig:fb_IC_capacity}). Our $n$-phase
scheme can achieve the two corner points within $2+3c_{JG}$ bits
per channel use for each user. The proof is using logarithmic Jensen's
gap property and is deferred to \ifarxiv  Appendix \ref{app:nphase_capacity}\else  \cite[Appendix K]{Joyson_jensens_gap_arxiv}\fi.

\subsection{An auxiliary result: Approximate capacity of 2-tap fast Fading ISI
channel}

Consider the 2-tap fast fading ISI channel described by
\begin{equation}
Y\brac l=g_{d}\brac lX\brac l+g_{c}\brac lX\brac{l-1}+Z\brac l,\label{eq:isi_channel}
\end{equation}
where $g_{d}\sim\mathcal{CN}\brac{0,SNR}$ and $g_{d}\sim\mathcal{CN}\brac{0,INR}$
are independent fading known only to the receiver and $Z\sim\mathcal{CN}\brac{0,1}$.
Also we assume a power constraint of $\expect{\abs X^{2}}\leq1$ on
the transmit symbols. Our analysis for $R_{1}$ can be easily modified
to obtain a closed form approximate expression for this channel. This
gives rise to the following corollary on the capacity of fading ISI
channels.
\begin{cor}
The capacity $C_{F-ISI}$ of the 2-tap fast fading ISI channel is
bounded by
\[
\lgbrac{1+SNR+INR}+1\geq C_{F-ISI}\geq\lgbrac{1+SNR+INR}-1-3c_{JG},
\]
where the channel fading strengths is assumed to have a logarithmic
Jensen's gap of $c_{JG}$.
\end{cor}
\begin{IEEEproof}
The proof is given in \ifarxiv  Appendix \ref{app:fading_isi_channel}\else  \cite[Appendix L]{Joyson_jensens_gap_arxiv}\fi.
\end{IEEEproof}

\section{Conclusion}

We introduced the notion of logarithmic Jensen's gap and demonstrated
that it can be used to obtain approximate capacity region for FF-ICs.
We proved that the rate-splitting schemes for ICs \cite{etkin_tse_no_fb_IC,chong2008han,suh_tse_fb_gaussian},
when extended to the fast fading case give capacity gap as a function
of the logarithmic Jensen's gap. Our analysis of logarithmic Jensen's
gap for fading models like Rayleigh fading show that rate-splitting
is approximately optimal for such cases. We then developed a scheme
for symmetric FF-ICs, which can be implemented using point-to-point
codes and can approximately achieve the capacity region. An important
direction to study will be to see if similar schemes with point-to-point
codes can be extended to general FF-ICs. Also our schemes are not
approximately optimal for bursty IC since it does not have finite
logarithmic Jensen's gap, it would be interesting to study if the
schemes can be extended to bursty IC and then to any arbitrary fading
distribution. Extension to FF-ICs with more than 2 users seems difficult,
since there are no approximate (within constant additive gap) capacity
results known even for 3-user IC with fixed channels.

\appendices{}

\section{Proof of achievability for non-feedback case\label{app:achievability_no_fb}}

We evaluate the term in the first inner bound inequality \eqref{eq:ach_nofb1}.
The other terms can be similarly evaluated.
\begin{align}
I\brac{X_{1};Y_{1},\underline{g_{1}}|U_{2}} & \overset{(a)}{=}I\brac{\rline{X_{1};Y_{1}}U_{2},\underline{g_{1}}}\\
 & =h\brac{\rline{Y_{1}}U_{2},\underline{g_{1}}}-h\brac{\rline{Y_{1}}X_{1},U_{2},\underline{g_{1}}},\\
h\brac{\rline{Y_{1}}U_{2},\underline{g_{1}}} & =h\brac{\rline{g_{11}X_{1}+g_{21}X_{2}+Z_{1}}U_{2},\underline{g_{1}}}\\
 & =h\brac{\rline{g_{11}X_{1}+g_{21}X_{p2}+Z_{1}}\underline{g_{1}}},\\
\text{variance}\brac{\rline{g_{11}X_{1}+g_{21}X_{p2}+Z_{1}}\underline{g_{1}}} & =\abs{g_{11}}^{2}+\lambda_{p2}\abs{g_{21}}^{2}+1,\\
\therefore h\brac{\rline{Y_{1}}U_{2},\underline{g_{1}}} & =\expect{\lgbrac{\abs{g_{11}}^{2}+\lambda_{p2}\abs{g_{21}}^{2}+1}}+\lgbrac{2\pi e},\\
h\brac{Y_{1}|X_{1},U_{2},\underline{g_{1}}} & =h\brac{\rline{g_{11}X_{1}+g_{21}X_{2}+Z_{1}}X_{1},U_{2},\underline{g_{1}}}\\
 & =h\brac{\rline{g_{21}X_{p2}+Z_{1}}\underline{g_{1}}}\\
 & =\expect{\lgbrac{1+\lambda_{p2}\abs{g_{21}}^{2}}}+\lgbrac{2\pi e}\\
 & \overset{(b)}{\leq}\expect{\lgbrac{1+\frac{1}{INR_{2}}\abs{g_{21}}^{2}}}+\lgbrac{2\pi e}\\
 & \overset{(c)}{\leq}\lgbrac 2+\lgbrac{2\pi e}\\
 & =1+\lgbrac{2\pi e},\\
\therefore I\brac{X_{1};Y_{1},\underline{g_{1}}|U_{2}} & \geq\expect{\lgbrac{\abs{g_{11}}^{2}+\lambda_{p2}\abs{g_{21}}^{2}+1}}-1,
\end{align}
where $(a)$ uses independence, $(b)$ is because $\lambda_{pi}\leq\frac{1}{INR_{i}}$,
and $(c)$ follows from Jensen's inequality.

\section{Proof of outer bounds for non-feedback case \label{app:outerbounds_nofb}}

Note that we have the notation $\underline{g}=[g_{11},g_{21},g_{22},g_{12}]$,
$S_{1}=g_{12}X_{1}+Z_{2},$ and $S_{2}=g_{21}X_{2}+Z_{1}.$ Our outer
bounding steps are valid while allowing $X_{1i}$ to be a function
of $\left(W_{1},\underline{g^{n}}\right)$, thus letting transmitters
have instantaneous and future CSIT. On choosing a uniform distribution
of messages we get
\begin{align}
 & n(R_{1}+2R_{2}-\epsilon_{n})\nonumber \\
 & \leq I\brac{W_{1};Y_{1}^{n},S_{1}^{n},\underline{g^{n}}}+I\brac{W_{2};Y_{2}^{n},\underline{g^{n}}}+I\brac{W_{2};Y_{2}^{n},S_{2}^{n},X_{1}^{n},\underline{g^{n}}}\\
 & =I\brac{\rline{W_{1};Y_{1}^{n},S_{1}^{n}}\underline{g^{n}}}+I\brac{\rline{W_{2};Y_{2}^{n}}\underline{g^{n}}}+I\brac{\rline{W_{2};Y_{2}^{n},S_{2}^{n}}X_{1}^{n},\underline{g^{n}}}\\
 & =I\brac{\rline{W_{1};S_{1}^{n}}\underline{g^{n}}}+I\brac{\rline{W_{1};Y_{1}^{n}}S_{1}^{n},\underline{g^{n}}}+I\brac{\rline{W_{2};Y_{2}^{n}}\underline{g^{n}}}\nonumber \\
 & \qquad+I\brac{\rline{W_{2};S_{2}^{n}}X_{1}^{n},\underline{g^{n}}}+I\brac{\rline{W_{2};Y_{2}^{n}}X_{1}^{n},S_{2}^{n},\underline{g^{n}}}\\
 & =h\brac{\rline{S_{1}^{n}}\underline{g^{n}}}-h\brac{\rline{S_{1}^{n}}W_{1},\underline{g^{n}}}+h\brac{\rline{Y_{1}^{n}}S_{1}^{n},\underline{g^{n}}}-h\brac{\rline{Y_{1}^{n}}W_{1},S_{1}^{n},\underline{g^{n}}}\nonumber \\
 & \qquad+h\brac{\rline{Y_{2}^{n}}\underline{g^{n}}}-h\brac{\rline{Y_{2}^{n}}W_{2},\underline{g^{n}}}+h\brac{\rline{S_{2}^{n}}X_{1}^{n},\underline{g^{n}}}-h\brac{\rline{S_{2}^{n}}X_{1}^{n},W_{2},\underline{g^{n}}}\nonumber \\
 & \qquad+h\brac{\rline{Y_{2}^{n}}X_{1}^{n},S_{2}^{n},\underline{g^{n}}}-h\brac{\rline{Y_{2}^{n}}X_{1}^{n},W_{2},S_{2}^{n},\underline{g^{n}}}\\
 & =h\brac{\rline{S_{1}^{n}}\underline{g^{n}}}-h\brac{Z_{2}^{n}}+h\brac{\rline{Y_{1}^{n}}S_{1}^{n},\underline{g^{n}}}-h\brac{\rline{S_{2}^{n}}\underline{g^{n}}}+h\brac{\rline{Y_{2}^{n}}\underline{g^{n}}}-h\brac{\rline{S_{1}^{n}}\underline{g^{n}}}\nonumber \\
 & \qquad+h\brac{\rline{S_{2}^{n}}\underline{g^{n}}}-h\brac{Z_{1}^{n}}+h\brac{\rline{Y_{2}^{n}}X_{1}^{n},S_{2}^{n},\underline{g^{n}}}-h\brac{Z_{2}^{n}}\\
 & =h\brac{\rline{Y_{1}^{n}}S_{1}^{n},\underline{g^{n}}}+h\brac{\rline{Y_{2}^{n}}\underline{g^{n}}}+h\brac{\rline{Y_{2}^{n}}X_{1}^{n},S_{2}^{n},\underline{g^{n}}}-h\brac{Z_{1}^{n}}-2h\brac{Z_{2}^{n}}\\
 & \overset{(a)}{\leq}\sum\sbrac{h\brac{Y_{1i}|S_{1i},\underline{g^{n}}}-h\brac{Z_{1i}}}+\sum\sbrac{h\brac{Y_{2i}|\underline{g^{n}}}-h\brac{Z_{2i}}}\nonumber \\
 & \qquad+\sum\sbrac{h\brac{Y_{2i}|X_{1i},S_{2i},\underline{g^{n}}}-h\brac{Z_{2i}}}\\
 & \overset{}{=}\mathbb{E}_{\underline{g^{n}}}\sbrac{\sum\brac{h\brac{Y_{1i}|S_{1i},\underline{g^{n}}}-h\brac{Z_{1i}}}}+\mathbb{E}_{\underline{g^{n}}}\sbrac{\sum\brac{h\brac{Y_{2i}|\underline{g^{n}}}-h\brac{Z_{2i}}}}\nonumber \\
 & \qquad+\mathbb{E}_{\underline{g^{n}}}\sbrac{\sum\brac{h\brac{Y_{2i}|X_{1i},S_{2i},\underline{g^{n}}}-h\brac{Z_{2i}}}}\\
 & \overset{(b)}{\leq}n\expect{\lgbrac{1+\abs{g_{21}}^{2}+\frac{\abs{g_{11}}^{2}}{1+\abs{g_{12}}^{2}}}}+n\expect{\lgbrac{1+\abs{g_{12}}^{2}+\abs{g_{22}}^{2}}}\nonumber \\
 & \qquad+n\expect{\lgbrac{1+\frac{\abs{g_{22}}^{2}}{1+\abs{g_{21}}^{2}}}},
\end{align}
where $\brac a$ is due to the fact that conditioning reduces entropy
and $\brac b$ follows from Equations \cite[(50)]{etkin_tse_no_fb_IC}
, \cite[(51)]{etkin_tse_no_fb_IC} and \cite[(52)]{etkin_tse_no_fb_IC}.
Note that in the calculation of step $\brac b$ we allow the symbols
$X_{1i},X_{2i}$ to depend on $\underline{g^{n}}$, but since $\underline{g^{n}}$
is available in conditioning the calculation proceeds similar to that
in \cite{etkin_tse_no_fb_IC}.
\begin{align}
 & n(R_{1}+R_{2}-\epsilon_{n})\\
 & \leq I\brac{W_{1};Y_{1}^{n},\underline{g^{n}}}+I\brac{W_{2};Y_{2}^{n},S_{2}^{n},X_{1}^{n},\underline{g^{n}}}\\
 & =I\brac{\rline{W_{1};Y_{1}^{n}}\underline{g^{n}}}+I\brac{\rline{W_{2};Y_{2}^{n},S_{2}^{n}}X_{1}^{n},\underline{g^{n}}}\\
 & =I\brac{\rline{W_{1};Y_{1}^{n}}\underline{g^{n}}}+I\brac{\rline{W_{2};S_{2}^{n}}X_{1}^{n},\underline{g^{n}}}+I\brac{\rline{W_{2};Y_{2}^{n}}X_{1}^{n},S_{2}^{n},\underline{g^{n}}}\\
 & =h\brac{\rline{Y_{1}^{n}}\underline{g^{n}}}-h\brac{\rline{Y_{1}^{n}}W_{1},\underline{g^{n}}}+h\brac{\rline{S_{2}^{n}}X_{1}^{n},\underline{g^{n}}}-h\brac{\rline{S_{2}^{n}}X_{1}^{n},W_{2},\underline{g^{n}}}\nonumber \\
 & \qquad+h\brac{\rline{Y_{2}^{n}}X_{1}^{n},S_{2}^{n},\underline{g^{n}}}-h\brac{\rline{Y_{2}^{n}}X_{1}^{n},W_{2},S_{2}^{n},\underline{g^{n}}}\\
 & =h\brac{\rline{Y_{1}^{n}}\underline{g^{n}}}-h\brac{\rline{S_{2}^{n}}\underline{g^{n}}}+h\brac{\rline{S_{2}^{n}}\underline{g^{n}}}-h\brac{Z_{1}^{n}}+h\brac{\rline{Y_{2}^{n}}X_{1}^{n},S_{2}^{n},\underline{g^{n}}}-h\brac{Z_{2}^{n}}\\
 & =h\brac{\rline{Y_{1}^{n}}\underline{g^{n}}}+h\brac{\rline{Y_{2}^{n}}X_{1}^{n},S_{2}^{n},\underline{g^{n}}}-h\brac{Z_{1}^{n}}-h\brac{Z_{2}^{n}}\\
 & \overset{(a)}{\leq}\sum\sbrac{h\brac{Y_{1i}|\underline{g^{n}}}-h\brac{Z_{1i}}}+\sum\sbrac{h\brac{Y_{2i}|X_{1i},S_{2i},\underline{g^{n}}}-h\brac{Z_{2i}}}\\
 & \overset{}{=}\mathbb{E}_{\underline{g^{n}}}\sbrac{\sum\brac{h\brac{Y_{1i}|\underline{g^{n}}}-h\brac{Z_{1i}}}}+\mathbb{E}_{\underline{g^{n}}}\sbrac{\sum\brac{h\brac{Y_{2i}|X_{1i},S_{2i},\underline{g^{n}}}-h\brac{Z_{2i}}}}\\
 & \overset{(b)}{\leq}n\expect{\lgbrac{1+\abs{g_{21}}^{2}+\abs{g_{11}}^{2}}}+n\expect{\lgbrac{1+\frac{\abs{g_{22}}^{2}}{1+\abs{g_{21}}^{2}}}},
\end{align}
where $\brac a$ is due to the fact that conditioning reduces entropy
and $\brac b$ again follows from Equations \cite[(51)]{etkin_tse_no_fb_IC}
and \cite[(52)]{etkin_tse_no_fb_IC}.

\bibliographystyle{ieeetr}
\bibliography{references,bibJournalList}

\clearpage

\section{Proof of Lemma \ref{lem:other_distr}\label{app:proof_bounded_log_lemma}}

We have $F\brac w\leq aw^{b}$ for $w\in[0,\epsilon]$, where $a\geq0,b>0,1\geq\epsilon>0$.
Now using integration by parts we get
\begin{align}
\expect{\lnbrac W} & \geq\int_{0}^{1}f\brac w\lnbrac wdw\\
 & =\int_{0}^{\epsilon}f\brac w\lnbrac wdw+\int_{\epsilon}^{1}f\brac w\lnbrac wdw\\
 & =\sbrac{F\brac w\lnbrac w}_{0}^{\epsilon}-\int_{0}^{\epsilon}F\brac w\frac{1}{w}dw+\int_{\epsilon}^{1}f\brac w\lnbrac wdw\\
 & \geq\sbrac{aw^{b}\lnbrac w}_{0}^{\epsilon}-\int_{0}^{\epsilon}aw^{b}\frac{1}{w}dw+\lnbrac{\epsilon}\\
 & \geq a\epsilon^{b}\lnbrac{\epsilon}-\frac{a\epsilon^{b}}{b}+\lnbrac{\epsilon}.
\end{align}
Note that $\lnbrac w$ is negative in the range $[0,1)$, thus we
get the desired inequalities in the last two steps.

\section{Proof of Corollary \ref{cor:fading=00003Dstatic_nofb}\label{app:proof_cor_fading_is_static_nofb}}

The rate region of non-feedback case in given in Equation \eqref{eq:inner_nofb}
can be reduced to the rate region for a channel without fading. Let
$\mathcal{R}'_{NFB}$ be the approximately optimal Han-Kobayashi rate
region of IC \cite{etkin_tse_no_fb_IC} with equivalent channel strengths
$SNR_{i}:=\expect{\abs{g_{ii}}^{2}}$ for $i=1,2$, and $INR_{i}:=\expect{\abs{g_{ij}}^{2}}$
for $i\neq j$. Then for a constant $c'$ we have
\begin{equation}
\mathcal{R}'_{NFB}\supseteq\mathcal{R}_{NFB}\supseteq\mathcal{R}'_{NFB}-c'.
\end{equation}
This can be verified by proceeding through each inner bound equation.
For example, consider the first inner bound Equation \eqref{eq:inner_nofb1}
$R_{1}\leq\expect{\lgbrac{1+\abs{g_{11}}^{2}+\lambda_{p2}\abs{g_{21}}^{2}}}-1.$
The corresponding equation in $\mathcal{R}'_{NFB}$ is $R_{1}\leq\lgbrac{1+SNR_{1}+\lambda_{p2}INR_{1}}-1.$
Now
\begin{align}
\lgbrac{1+SNR_{1}+\lambda_{p2}INR_{1}}-1 & \overset{\brac a}{\geq}\expect{\lgbrac{1+\abs{g_{11}}^{2}+\lambda_{p2}\abs{g_{21}}^{2}}}-1\label{eq:fading=00003Dstatic_nofb1}\\
 & \overset{\brac b}{\geq}\brac{\lgbrac{1+SNR_{1}+\lambda_{p2}INR_{1}}-1}-2c_{JG},\label{eq:fading=00003Dstatic_nofb2}
\end{align}
where $\brac a$ is due to Jensen's inequality and $\brac b$ is using
logarithmic Jensen's gap result twice. Due to \eqref{eq:fading=00003Dstatic_nofb1},
\eqref{eq:fading=00003Dstatic_nofb2} it follows that the first inner
bound equation for fading case is in constant gap with that of static
case. Similarly, by proceeding through each inner bound equation,
it follows that $\mathcal{R}'_{NFB}\supseteq\mathcal{R}_{NFB}\supseteq\mathcal{R}'_{NFB}-c'$
for a constant $c'$.

\section{Proof of Theorem \ref{thm:IMA} (Fast fading interference multiple
access channel)\label{app:IMA}}

We have the following achievable rate region for from fast fading
interference multiple access channel, by using the scheme from \cite{perron_IMA}
by considering $\brac{Y_{1},\underline{g_{1}}}$, $\brac{Y_{2},\underline{g_{2}}}$
as the outputs at the receivers.
\begin{align}
R_{1} & \leq I\brac{X_{1};Y_{1},\underline{g_{1}}|X_{2}}\label{eq:ach_IMA1}\\
R_{2} & \leq I\brac{X_{2};Y_{2},\underline{g_{2}}|U_{1}}\label{eq:ach_IMA2}\\
R_{2} & \leq I\brac{X_{2};Y_{1},\underline{g_{2}}|X_{1}}\label{eq:ach_IMA3}\\
R_{1}+R_{2} & \leq I\brac{X_{1}X_{2};Y_{1},\underline{g_{1}}}\label{eq:ach_IMA4}\\
R_{1}+R_{2} & \leq I\brac{X_{2},U_{1};Y_{2},\underline{g_{2}}}+I\brac{X_{1};Y_{1},\underline{g_{1}}|U_{1},X_{2}}\label{eq:ach_IMA5}\\
R_{1}+2R_{2} & \leq I\brac{X_{2},U_{1};Y_{2},\underline{g_{2}}}+I\brac{X_{1},X_{2};Y_{1},\underline{g_{1}}|U_{1}}\label{eq:ach_IMA6}
\end{align}
with mutually independent Gaussian input distributions $U_{1},X_{p1},X_{2}$
\begin{equation}
U_{1}\sim\mathcal{CN}\brac{0,\lambda_{c1}},\quad X_{p1}\sim\mathcal{CN}\brac{0,\lambda_{p1}},
\end{equation}
\begin{equation}
X_{1}=U_{1}+X_{p1},\quad X_{2}\sim\mathcal{CN}\brac{0,1},
\end{equation}
where $\lambda_{c1}+\lambda_{p1}=1$ and $\lambda_{p1}=\min\brac{\frac{1}{INR_{1}},1}$.
Evaluating the achievable region we obtain
\begin{align}
R_{1} & \leq\expect{\lgbrac{1+\abs{g_{11}}^{2}}}\label{eq:inner_IMA1}\\
R_{2} & \leq\expect{\lgbrac{1+\abs{g_{22}}^{2}+\lambda_{p1}\abs{g_{12}}^{2}}}-1\label{eq:inner_IMA2}\\
R_{2} & \leq\expect{\lgbrac{1+\abs{g_{21}}^{2}}}\label{eq:inner_IMA3}\\
R_{1}+R_{2} & \leq\expect{\lgbrac{1+\abs{g_{11}}^{2}+\abs{g_{21}}^{2}}}\label{eq:inner_IMA4}\\
R_{1}+R_{2} & \leq\expect{\lgbrac{1+\abs{g_{22}}^{2}+\abs{g_{12}}^{2}}}+\expect{\lgbrac{1+\lambda_{p1}\abs{g_{11}}^{2}}}-1\label{eq:inner_IMA5}\\
R_{1}+2R_{2} & \leq\expect{\lgbrac{1+\abs{g_{22}}^{2}+\abs{g_{12}}^{2}}}+\expect{\lgbrac{1+\lambda_{p1}\abs{g_{11}}^{2}+\abs{g_{21}}^{2}}}-1.\label{eq:inner_IMA6}
\end{align}
The calculations are similar to that of FF-IC (subsection \vref{subsec:no_fb}).
Now we claim the following outer bounds.
\begin{align}
R_{1} & \leq\expect{\lgbrac{1+\abs{g_{11}}^{2}}}\label{eq:outer_IMA1}\\
R_{2} & \leq\expect{\lgbrac{1+\abs{g_{22}}^{2}}}\label{eq:outer_IMA2}\\
R_{2} & \leq\expect{\lgbrac{1+\abs{g_{21}}^{2}}}\label{eq:outer_IMA3}\\
R_{1}+R_{2} & \leq\expect{\lgbrac{1+\abs{g_{11}}^{2}+\abs{g_{21}}^{2}}}\label{eq:outer_IMA4}\\
R_{1}+R_{2} & \leq\expect{\lgbrac{1+\abs{g_{22}}^{2}+\abs{g_{12}}^{2}}}+\expect{\lgbrac{1+\frac{\abs{g_{11}}^{2}}{1+\abs{g_{12}}^{2}}}}\label{eq:outer_IMA5}\\
R_{1}+2R_{2} & \leq\expect{\lgbrac{1+\abs{g_{22}}^{2}+\abs{g_{12}}^{2}}}+\expect{\lgbrac{1+\frac{\abs{g_{11}}^{2}}{1+\abs{g_{12}}^{2}}+\abs{g_{21}}^{2}}}.\label{eq:outer_IMA6}
\end{align}

With the above outer bounds it can be shown that the capacity region
can be achieved within $1+\frac{1}{2}c_{JG}$ bits per channel use.
The computations are similar to that of FF-IC (claim \vref{claim:capacity_gap_noFB}).

The outer bound \eqref{eq:outer_IMA1} can be derived by giving side
information $W_{2}$ at Rx1 and requiring $W_{1}$ to be decoded,
outer bound \eqref{eq:outer_IMA2} can be derived by giving side information
$W_{1}$ at Rx2 and requiring $W_{2}$ to be decoded, outer bound
\eqref{eq:outer_IMA3} can be derived by giving side information $W_{1}$
at Rx1 and requiring $W_{2}$ to be decoded, outer bound \eqref{eq:outer_IMA4}
can be derived by giving side information requiring $W_{1},W_{2}$
to be decoded at Rx1 with no side information. We derive \eqref{eq:outer_IMA5}
below. Note that we have the notation $\underline{g}=[g_{11},g_{21},g_{22},g_{12}]$,
$S_{1}=g_{12}X_{1}+Z_{2},$ and $S_{2}=g_{21}X_{2}+Z_{1}.$
\begin{align}
 & n(R_{2}+R_{1}-\epsilon_{n})\\
 & \leq I\brac{W_{2};Y_{2}^{n},\underline{g^{n}}}+I\brac{W_{1};Y_{1}^{n},S_{1}^{n},X_{2}^{n},\underline{g^{n}}}\\
 & =I\brac{\rline{W_{2};Y_{2}^{n}}\underline{g^{n}}}+I\brac{\rline{W_{1};Y_{1}^{n},S_{1}^{n}}X_{2}^{n},\underline{g^{n}}}\\
 & =I\brac{\rline{W_{2};Y_{2}^{n}}\underline{g^{n}}}+I\brac{\rline{W_{1};S_{1}^{n}}X_{2}^{n},\underline{g^{n}}}+I\brac{\rline{W_{1};Y_{1}^{n}}X_{2}^{n},S_{1}^{n},\underline{g^{n}}}\\
 & =h\brac{\rline{Y_{2}^{n}}\underline{g^{n}}}-h\brac{\rline{Y_{2}^{n}}W_{2},\underline{g^{n}}}+h\brac{\rline{S_{1}^{n}}X_{2}^{n},\underline{g^{n}}}-h\brac{\rline{S_{1}^{n}}X_{2}^{n},W_{1},\underline{g^{n}}}\nonumber \\
 & \qquad+h\brac{\rline{Y_{1}^{n}}X_{2}^{n},S_{1}^{n},\underline{g^{n}}}-h\brac{\rline{Y_{1}^{n}}X_{2}^{n},W_{1},S_{1}^{n},\underline{g^{n}}}\\
 & =h\brac{\rline{Y_{2}^{n}}\underline{g^{n}}}-h\brac{\rline{S_{1}^{n}}\underline{g^{n}}}+h\brac{\rline{S_{1}^{n}}\underline{g^{n}}}-h\brac{Z_{2}^{n}}+h\brac{\rline{Y_{1}^{n}}X_{2}^{n},S_{1}^{n},\underline{g^{n}}}-h\brac{Z_{1}^{n}}\\
 & =h\brac{\rline{Y_{2}^{n}}\underline{g^{n}}}+h\brac{\rline{Y_{1}^{n}}X_{2}^{n},S_{1}^{n},\underline{g^{n}}}-h\brac{Z_{2}^{n}}-h\brac{Z_{1}^{n}}\\
 & \overset{(b)}{\leq}\sum\sbrac{h\brac{Y_{2i}|\underline{g^{n}}}-h\brac{Z_{2i}}}+\sum\sbrac{h\brac{Y_{1i}|X_{2i},S_{1i},\underline{g^{n}}}-h\brac{Z_{1i}}}\\
 & \overset{}{=}\mathbb{E}_{\underline{g^{n}}}\sbrac{\sum\brac{h\brac{Y_{2i}|\underline{g^{n}}}-h\brac{Z_{2i}}}}+\mathbb{E}_{\underline{g^{n}}}\sbrac{\sum\brac{h\brac{Y_{1i}|X_{2i},S_{1i},\underline{g^{n}}}-h\brac{Z_{1i}}}}\\
 & \overset{(c)}{\leq}n\expect{\lgbrac{1+\abs{g_{12}}^{2}+\abs{g_{22}}^{2}}}+n\expect{\lgbrac{1+\frac{\abs{g_{11}}^{2}}{1+\abs{g_{12}}^{2}}}},
\end{align}
where $\brac a$ is because $\brac{X_{2}^{n},\underline{g^{n}}}$
is independent of $W_{1}$, $\brac b$ is due to the fact that conditioning
reduces entropy and $\brac c$ follows from Equations \cite[(51)]{etkin_tse_no_fb_IC}
and \cite[(52)]{etkin_tse_no_fb_IC}. Now we derive \eqref{eq:outer_IMA6}
below using the fact that $W_{2}$ has to be decoded at both receivers:
\begin{align}
 & n(R_{1}+2R_{2}-\epsilon_{n})\nonumber \\
 & \leq I\brac{W_{1};Y_{1}^{n},S_{1}^{n},\underline{g^{n}}}+I\brac{W_{2};Y_{2}^{n},\underline{g^{n}}}+I\brac{W_{2};Y_{1}^{n},S_{2}^{n},\underline{g^{n}}}\\
 & =I\brac{\rline{W_{1};Y_{1}^{n},S_{1}^{n}}\underline{g^{n}}}+I\brac{\rline{W_{2};Y_{2}^{n}}\underline{g^{n}}}+I\brac{\rline{W_{2};Y_{1}^{n},S_{2}^{n}}\underline{g^{n}}}\\
 & =I\brac{\rline{W_{1};S_{1}^{n}}\underline{g^{n}}}+I\brac{\rline{W_{1};Y_{1}^{n}}S_{1}^{n},\underline{g^{n}}}+I\brac{\rline{W_{2};Y_{2}^{n}}\underline{g^{n}}}\nonumber \\
 & \qquad+I\brac{\rline{W_{2};S_{2}^{n}}\underline{g^{n}}}+\underset{=0}{\underbrace{I\brac{\rline{W_{2};Y_{1}^{n}}S_{2}^{n},\underline{g^{n}}}}}\\
 & =h\brac{\rline{S_{1}^{n}}\underline{g^{n}}}-h\brac{\rline{S_{1}^{n}}W_{1},\underline{g^{n}}}+h\brac{\rline{Y_{1}^{n}}S_{1}^{n},\underline{g^{n}}}-h\brac{\rline{Y_{1}^{n}}W_{1},S_{1}^{n},\underline{g^{n}}}\nonumber \\
 & \qquad+h\brac{\rline{Y_{2}^{n}}\underline{g^{n}}}-h\brac{\rline{Y_{2}^{n}}W_{2},\underline{g^{n}}}+h\brac{\rline{S_{2}^{n}}\underline{g^{n}}}-h\brac{\rline{S_{2}^{n}}W_{2},\underline{g^{n}}}\\
 & =h\brac{\rline{S_{1}^{n}}\underline{g^{n}}}-h\brac{Z_{2}^{n}}+h\brac{\rline{Y_{1}^{n}}S_{1}^{n},\underline{g^{n}}}-h\brac{\rline{S_{2}^{n}}\underline{g^{n}}}+h\brac{\rline{Y_{2}^{n}}\underline{g^{n}}}-h\brac{\rline{S_{1}^{n}}\underline{g^{n}}}\nonumber \\
 & \qquad+h\brac{\rline{S_{2}^{n}}\underline{g^{n}}}-h\brac{Z_{1}^{n}}\\
 & =h\brac{\rline{Y_{1}^{n}}S_{1}^{n},\underline{g^{n}}}+h\brac{\rline{Y_{2}^{n}}\underline{g^{n}}}-h\brac{Z_{1}^{n}}-h\brac{Z_{2}^{n}}\\
 & \overset{(a)}{\leq}\sum\sbrac{h\brac{Y_{1i}|S_{1i},\underline{g^{n}}}-h\brac{Z_{1i}}}+\sum\sbrac{h\brac{Y_{2i}|\underline{g^{n}}}-h\brac{Z_{2i}}}\\
 & \overset{}{=}\mathbb{E}_{\underline{g^{n}}}\sbrac{\sum\brac{h\brac{Y_{1i}|S_{1i},\underline{g^{n}}}-h\brac{Z_{1i}}}}+\mathbb{E}_{\underline{g^{n}}}\sbrac{\sum\brac{h\brac{Y_{2i}|\underline{g^{n}}}-h\brac{Z_{2i}}}}\\
 & \overset{(b)}{\leq}n\expect{\lgbrac{1+\abs{g_{21}}^{2}+\frac{\abs{g_{11}}^{2}}{1+\abs{g_{12}}^{2}}}}+n\expect{\lgbrac{1+\abs{g_{12}}^{2}+\abs{g_{22}}^{2}}},
\end{align}
where $\brac a$ is due to the fact that conditioning reduces entropy
and $\brac b$ follows from Equations \cite[(50)]{etkin_tse_no_fb_IC}
, \cite[(51)]{etkin_tse_no_fb_IC} and \cite[(52)]{etkin_tse_no_fb_IC}.
Note that in the calculation of step $\brac b$ we allow the symbols
$X_{1i},X_{2i}$ to depend on $\underline{g^{n}}$, but since $\underline{g^{n}}$
is available in conditioning the calculation proceeds similar to that
in \cite{etkin_tse_no_fb_IC}.

\section{Proof of Corollary \ref{cor:static=00003Dfading_fb} \label{app:proof_cor_fading_is_static_fB}}

Let $\mathcal{R}'_{NFB}$ be the approximately optimal Han-Kobayashi
rate region of feedback IC \cite{suh_tse_fb_gaussian} with equivalent
channel strengths $SNR_{i}:=\expect{\abs{g_{ii}}^{2}}$ for $i=1,2$,
and $INR_{i}:=\expect{\abs{g_{ij}}^{2}}$ for $i\neq j$. Then for
a constant $c''$ we have
\begin{equation}
\mathcal{R}'_{FB}\supseteq\mathcal{R}_{FB}\supseteq\mathcal{R}'_{FB}-c''.
\end{equation}
This can be verified by proceeding through each inner bound equation.
For example, consider the first inner bound Equation \eqref{eq:inner_fb1}
$R_{1}\leq\expect{\lgbrac{\abs{g_{11}}^{2}+\abs{g_{21}}^{2}+2\abs{\rho}^{2}\text{Re}\brac{g_{11}g_{21}^{*}}+1}}-1.$
The corresponding equation in $\mathcal{R}'_{NFB}$ is $R_{1}\leq\lgbrac{1+SNR_{1}+INR_{2}+2\abs{\rho}^{2}\sqrt{SNR_{1}\cdot INR_{2}}+1}-1.$%
{} Now
\begin{align}
 & \expect{\lgbrac{\abs{g_{11}}^{2}+\abs{g_{21}}^{2}+2\abs{\rho}^{2}\text{Re}\brac{g_{11}g_{21}^{*}}+1}}\nonumber \\
 & \overset{\brac a}{\leq}\lgbrac{1+SNR_{1}+INR_{2}}\\
 & \leq\lgbrac{1+SNR_{1}+INR_{2}+2\abs{\rho}^{2}\sqrt{SNR_{1}\cdot INR_{2}}+1},\label{eq:fading=00003Dstatic_fb1}
\end{align}
where $\brac a$ is due to Jensen's inequality and independence of
$g_{11},g_{21}$. Also
\begin{align}
 & \expect{\lgbrac{\abs{g_{11}}^{2}+\abs{g_{21}}^{2}+2\abs{\rho}^{2}\text{Re}\brac{g_{11}g_{21}^{*}}+1}}\nonumber \\
 & \overset{\brac a}{=}\expect{\lgbrac{\abs{g_{11}}^{2}+\abs{g_{21}}^{2}+2\abs{g_{11}}\abs{g_{21}}\abs{\rho}\cos\brac{\theta}+1}}\\
 & \overset{\brac b}{\geq}\expect{\lgbrac{\abs{g_{11}}^{2}+\abs{g_{21}}^{2}+1}}-1\\
 & \overset{\brac c}{\geq}\lgbrac{SNR_{1}+INR_{2}+1}-1-2c_{JG}\\
 & \overset{\brac d}{\geq}\lgbrac{SNR_{1}+INR_{2}+2\abs{\rho}^{2}\sqrt{SNR_{1}\cdot INR_{2}}+1}-2-2c_{JG},\label{eq:fading=00003Dstatic_fb2}
\end{align}
where $\brac a$ is because phases of $g_{11},g_{12}$ are independently
uniformly distributed in $\sbrac{0,2\pi}$ yielding $\text{Re}\brac{g_{11}g_{21}^{*}}=\abs{g_{11}}\abs{g_{21}}\cos\brac{\theta}$
with an independent $\theta\sim\text{Unif}\sbrac{0,2\pi}$, $\brac b$
is using the fact that for $p>q$ $\frac{1}{2\pi}\int_{0}^{2\pi}\lgbrac{p+q\cos\brac{\theta}}d\theta=\lgbrac{\frac{p+\sqrt{p^{2}-q^{2}}}{2}}\geq\lgbrac p-1$,
$\brac c$ is using logarithmic Jensen's gap result twice and $\brac d$
is because $SNR_{1}+INR_{2}\geq2\abs{\rho}^{2}\sqrt{SNR_{1}\cdot INR_{2}}$.
It follows from Equations \eqref{eq:fading=00003Dstatic_fb1} and
\eqref{eq:fading=00003Dstatic_fb2}, that the first inner bound for
fading case is within constant gap with the first inner bound of the
static case.

Now consider the second inner bound Equation \eqref{eq:inner_fb2}
\begin{equation}
R_{1}\leq\expect{\lgbrac{1+\brac{1-\abs{\rho}^{2}}\abs{g_{12}}^{2}}}+\expect{\lgbrac{1+\lambda_{p1}\abs{g_{11}}^{2}+\lambda_{p2}\abs{g_{21}}^{2}}}-2
\end{equation}
and the corresponding equation
\begin{equation}
R_{1}\leq\lgbrac{1+\brac{1-\abs{\rho}^{2}}INR_{1}}+\lgbrac{1+\lambda_{p1}SNR_{1}+\lambda_{p2}INR_{2}}-3c_{JG}-2
\end{equation}
from $\mathcal{R}'_{FB}$. We have
\begin{align}
 & \expect{\lgbrac{1+\brac{1-\abs{\rho}^{2}}\abs{g_{12}}^{2}}}+\expect{\lgbrac{1+\lambda_{p1}\abs{g_{11}}^{2}+\lambda_{p2}\abs{g_{21}}^{2}}}\nonumber \\
 & \leq\lgbrac{1+\brac{1-\abs{\rho}^{2}}INR_{1}}+\lgbrac{1+\lambda_{p1}SNR_{1}+\lambda_{p2}INR_{2}}\label{eq:fading=00003Dstatic_fb3}
\end{align}
due to Jensen's inequality. And
\begin{align}
 & \expect{\lgbrac{1+\brac{1-\abs{\rho}^{2}}\abs{g_{12}}^{2}}}+\expect{\lgbrac{1+\lambda_{p1}\abs{g_{11}}^{2}+\lambda_{p2}\abs{g_{21}}^{2}}}\nonumber \\
 & \geq\lgbrac{1+\brac{1-\abs{\rho}^{2}}INR_{1}}+\lgbrac{1+\lambda_{p1}SNR_{1}+\lambda_{p2}INR_{2}}-3c_{JG}\label{eq:fading=00003Dstatic_fb4}
\end{align}
using logarithmic Jensen's gap result thrice. It follows from Equations
\eqref{eq:fading=00003Dstatic_fb3} and \eqref{eq:fading=00003Dstatic_fb4},
that the second inner bound for fading case is within constant gap
with the second inner bound of the static case. Similarly, by proceeding
through each inner bound equation, it follows that
\[
\mathcal{R}'_{FB}\supseteq\mathcal{R}_{FB}\supseteq\mathcal{R}'_{FB}-c''
\]
for a constant $c''$.

\section{Proof of achievability for feedback case\label{app:achievability_fb}}

We evaluate the term in the first inner bound inequality $(\ref{eq:ach_fb1})$
. The other terms can be similarly evaluated.
\begin{align}
I\brac{U,U_{2},X_{1};Y_{1},\underline{g_{1}}} & \overset{(a)}{=}I\brac{\rline{U,U_{2},X_{1};Y_{1}}\underline{g_{1}}}\\
 & =h\brac{Y_{1}|\underline{g_{1}}}-h\brac{Y_{1}|\underline{g_{1}},U,U_{2},X_{1}},
\end{align}
\begin{align}
\text{variance}\brac{Y_{1}|\underline{g_{1}}} & =\text{variance}\brac{\rline{g_{11}X_{1}+g_{21}X_{2}+Z_{1}}g_{11},g_{21}}\\
 & =\abs{g_{11}}^{2}+\abs{g_{21}}^{2}+g_{11}^{*}g_{21}\expect{X_{1}^{*}X_{2}}+g_{11}g_{21}^{*}\expect{X_{1}X_{2}^{*}}+1\\
 & =\abs{g_{11}}^{2}+\abs{g_{21}}^{2}+2\abs{\rho}^{2}\text{Re}\brac{g_{11}g_{21}^{*}}+1,
\end{align}
\begin{align}
h\brac{Y_{1}|\underline{g_{1}},U,U_{2},X_{1}} & =h\brac{\rline{g_{11}X_{1}+g_{21}X_{2}+Z_{1}}\underline{g_{1}},U,U_{2},X_{1}}\\
 & =h\brac{\rline{g_{21}X_{p2}+Z_{1}}\underline{g_{1}}}\\
 & =\expect{\lgbrac{1+\lambda_{p2}\abs{g_{21}}^{2}}}+\lgbrac{2\pi e}\\
 & \overset{(b)}{\leq}\expect{\lgbrac{1+\frac{1}{INR_{2}}\abs{g_{21}}^{2}}}+\lgbrac{2\pi e}\\
 & \overset{(c)}{\leq}\lgbrac 2+\lgbrac{2\pi e}\\
 & =1+\lgbrac{2\pi e},\\
\therefore I\brac{U,U_{2},X_{1};Y_{1},\underline{g_{1}}} & \geq\expect{\lgbrac{\abs{g_{11}}^{2}+\abs{g_{21}}^{2}+2\abs{\rho}^{2}\text{Re}\brac{g_{11}g_{21}^{*}}+1}}-1,
\end{align}
where $(a)$ uses independence, $(b)$ is because $\lambda_{pi}\leq\frac{1}{INR_{i}}$,
and $(c)$ follows from Jensen's inequality.

\section{Proof of outer bounds for feedback case\label{app:outer_bounds_fb}}

Following the methods in \cite{suh_tse_fb_gaussian}, we let $\expect{X_{1}X_{2}^{*}}=\rho$.
We have the notation $\underline{g_{1}}=[g_{11},g_{21}]$ , $\underline{g_{2}}=[g_{22},g_{12}]$,
$\underline{g}=[g_{11},g_{21},g_{22},g_{12}]$, $S_{1}=g_{12}X_{1}+Z_{2},$
and $S_{2}=g_{21}X_{2}+Z_{1}.$ We let $\expect{X_{1}X_{2}^{*}}=\rho=\abs{\rho}e^{i\theta}$.
All of our outer bounding steps are valid while allowing $X_{1i}$
to be a function of $\left(W_{1},Y_{1}^{i-1},\underline{g_{1}^{n}}\right)$,
thus letting transmitters have full CSIT along with feedback. On choosing
a uniform distribution of messages we get
\begin{align}
n(R_{1}-\epsilon_{n}) & \overset{(a)}{\leq}I\brac{\rline{W_{1};Y_{1}^{n}}\underline{g_{1}^{n}}}\\
 & \overset{(b)}{\leq}\sum\brac{h\brac{Y_{1i}|\underline{g_{1i}}}-h\brac{Z_{1i}}}\\
 & =\sum\brac{\mathbb{E}_{\underline{\tilde{g}_{1i}}}\sbrac{h\brac{Y_{1i}|\underline{g_{1i}}=\underline{\tilde{g}_{1i}}}-h\brac{Z_{1i}}}}\\
 & \overset{(c)}{=}\mathbb{E}_{\underline{\tilde{g}_{1}}}\sbrac{\sum\brac{h\brac{Y_{1i}|\underline{g_{1i}}=\underline{\tilde{g}_{1}}}-h\brac{Z_{1i}}}}\\
\therefore R_{1} & \leq\expect{\lgbrac{\abs{g_{11}}^{2}+\abs{g_{21}}^{2}+\brac{\rho^{*}g_{11}^{*}g_{21}+\rho g_{11}g_{21}^{*}}+1}},
\end{align}
where $(a)$ follows from Fano's inequality, $(b)$ follows from the
fact that conditioning reduces entropy, and $(c)$ follows from the
fact that $\underline{\tilde{g}_{1i}}$ are i.i.d. Now we bound $R_{1}$
in a second way as done in \cite{suh_tse_fb_gaussian}:
\begin{align}
n(R_{1} & -\epsilon_{n})\leq I\brac{W_{1};Y_{1}^{n},\underline{g_{1}^{n}}}\\
 & \leq I\brac{W_{1};Y_{1}^{n},\underline{g_{1}^{n}},Y_{2}^{n},\underline{g_{2}^{n}},W_{2}}\\
 & =I\brac{W_{1};\underline{g}^{n},W_{2}}+I\brac{W_{1};Y_{1}^{n},Y_{2}^{n}|\underline{g}^{n},W_{2}}\\
 & =0+I\brac{W_{1};Y_{1}^{n},Y_{2}^{n}|\underline{g}^{n},W_{2}}\\
 & =h\brac{Y_{1}^{n},Y_{2}^{n}|\underline{g}^{n},W_{2}}-h\brac{Y_{1}^{n},Y_{2}^{n}|\underline{g}^{n},W_{1},W_{2}}\\
 & =\sum\sbrac{h\brac{Y_{1i},Y_{2i}|\underline{g}^{n},W_{2},Y_{1}^{i-1},Y_{2}^{i-1}}}-\sum\sbrac{h\brac{Z_{1i}}+h\brac{Z_{2i}}}\\
 & =\sum\sbrac{h\brac{Y_{2i}|\underline{g}^{n},W_{2},Y_{1}^{i-1},Y_{2}^{i-1}}}+\sum\sbrac{h\brac{Y_{1i}|\underline{g}^{n},W_{2},Y_{1}^{i-1},Y_{2}^{i}}}\nonumber \\
 & \qquad-\sum\sbrac{h\brac{Z_{1i}}+h\brac{Z_{2i}}}\\
 & \overset{(a)}{=}\sum\sbrac{h\brac{Y_{2i}|\underline{g}^{n},W_{2},Y_{1}^{i-1},Y_{2}^{i-1},X_{2}^{i}}}+\sum\sbrac{h\brac{Y_{1i}|\underline{g}^{n},W_{2},Y_{1}^{i-1},Y_{2}^{i},S_{1i},X_{2}^{i}}}\nonumber \\
 & \qquad-\sum\sbrac{h\brac{Z_{1i}}+h\brac{Z_{2i}}}\\
 & \overset{(b)}{\leq}\sum\sbrac{h\brac{Y_{2i}|\underline{g_{i}},X_{2i}}-h\brac{Z_{2i}}}+\sum\sbrac{h\brac{Y_{1i}|\underline{g_{i}},S_{1i},X_{2i}}-h\brac{Z_{1i}}}\\
 & \overset{(c)}{=}\mathbb{E}_{\underline{\tilde{g}}}\sbrac{\sum\brac{h\brac{Y_{2i}|X_{2i},\underline{g_{i}}=\underline{\tilde{g}}}-h\brac{Z_{2i}}}}\nonumber \\
 & \qquad+\mathbb{E}_{\underline{\tilde{g}}}\sbrac{\sum\brac{h\brac{Y_{1i}|S_{1i},X_{2i},\underline{g_{i}}=\underline{\tilde{g}}}-h\brac{Z_{1i}}}},
\end{align}
\begin{equation}
\therefore\begin{aligned}R_{1} & \overset{(d)}{\leq}\expect{\lgbrac{1+\brac{1-\abs{\rho}^{2}}\abs{g_{12}}^{2}}}+\expect{\lgbrac{1+\frac{\brac{1-\abs{\rho}^{2}}\abs{g_{11}}^{2}}{1+\brac{1-\abs{\rho}^{2}}\abs{g_{12}}^{2}}}}\end{aligned}
,
\end{equation}
where $(a)$ follows from the fact that $X_{2}^{i}$ is a function
of $\brac{W_{2},Y_{2}^{i-1},\underline{g}^{n}}$ and $S_{1i}$ is
a function of $\brac{Y_{2}^{i},X_{2}^{i},\underline{g}^{n}}$, $(b)$
follows from the fact that conditioning reduces entropy, $(c)$ follows
from the fact that $\underline{\tilde{g}_{i}}$ are i.i.d., and $(d)$
follows from \cite[(43)]{suh_tse_fb_gaussian}. The other outer bounds
can be derived similarly following \cite{suh_tse_fb_gaussian} and
making suitable changes to account for fading as we illustrated in
the previous two derivations.

\section{Fading matrix\label{app:Fadingmatrix}}

The calculations are given in Equations \eqref{eq:fading_det_calculation1},\eqref{eq:fading_det_calculation2}.{\small{}
}
\begin{align}
 & \expect{\lgbrac{\abs{K_{{\bf Y_{1}}}(n)}}}\nonumber \\
 & =\mathbb{E}\left[\log\left(\brac{\abs{g_{11}\brac n}^{2}+\abs{g_{21}\brac n}^{2}\brac{\frac{\abs{g_{12}\brac{n-1}}^{2}+1}{1+INR}}+1}\abs{K_{{\bf Y_{1}}}(n-1)}\right.\right.\nonumber \\
 & \qquad\qquad\left.\left.-\frac{\abs{g_{11}\brac{n-1}}^{2}\abs{g_{21}\brac n}^{2}\abs{g_{12}\brac{n-1}}^{2}}{1+INR}\abs{K_{{\bf Y_{1}}}(n-2)}\right)\right]\label{eq:fading_det_calculation1}\\
 & \geq\mathbb{E}\left[\log\left(\brac{1+INR+SNR}\abs{K_{{\bf Y_{1}}}(n-1)}\right.\right.\nonumber \\
 & \qquad\qquad\left.\left.-\frac{INR\cdot INR\abs{g_{11}\brac{n-1}}^{2}}{1+INR}\abs{K_{{\bf Y_{1}}}(n-2)}\right)\right]-3c_{JG}.\label{eq:fading_det_calculation2}
\end{align}
\sloppy The first step \eqref{eq:fading_det_calculation1}, is by
expanding the determinant. We use the logarithmic Jensen's gap property
thrice in the second step \eqref{eq:fading_det_calculation2}. This
is justified because the coefficients of $\cbrac{\abs{g_{11}\brac n}^{2},\abs{g_{12}\brac{n-1}}^{2},\abs{g_{21}\brac n}^{2}}$
from Equation \eqref{eq:fading_det_calculation1} are non-negative
(due to the fact that all the matrices involved are covariance matrices),
and the coefficients themselves are independent of $\cbrac{\abs{g_{11}\brac n}^{2},\abs{g_{12}\brac{n-1}}^{2},\abs{g_{21}\brac n}^{2}}$.
(Note that $\abs{K_{{\bf Y_{1}}}(n-1)}$ depend on $\abs{g_{12}\brac{n-2}}^{2}$
but not on $\abs{g_{12}\brac{n-1}}^{2}$). This procedure can be carried
out $n$ times and it follows that:
\begin{align}
\underset{n\rightarrow\infty}{\text{lim}}\frac{1}{n}\expect{\lgbrac{\abs{K_{{\bf Y_{1}}}(n)}}} & \geq\underset{n\rightarrow\infty}{\text{lim}}\frac{1}{n}\lgbrac{\abs{\hat{K}_{{\bf Y_{1}}}(n)}}-3c_{JG},
\end{align}
where $\hat{K}_{{\bf Y_{1}}}(n)$ is obtained from $K_{{\bf Y_{1}}}(n)$
by replacing $g_{12}\brac i$'s, $g_{21}\brac i$'s with $\sqrt{INR}$
and $g_{11}\brac i$'s with $\sqrt{SNR}$.

\section{Matrix determinant: asymptotic behavior \label{app:proof_matrix_asym}}

The following recursion easily follows:
\begin{equation}
\abs{A_{n}}=\abs a\abs{A_{n-1}}-\abs b^{2}\abs{A_{n-2}}
\end{equation}
with $\abs{A_{1}}=\abs a,\abs{A_{2}}=\abs a^{2}-\abs b^{2}$. Also
$\abs{A_{0}}$ can be consistently defined to be $1$. The characteristic
equation for this recursive relation is given by: $\lambda^{2}-\abs a\lambda+\abs b^{2}=0$
and the characteristic roots are given by:
\begin{equation}
\lambda_{1}=\frac{\abs a+\sqrt{\abs a^{2}-4\abs b^{2}}}{2},\lambda_{2}=\frac{\abs a-\sqrt{\abs a^{2}-4\abs b^{2}}}{2}.
\end{equation}
Now the solution of the recursive system is given by $\abs{A_{n}}=c_{1}\lambda_{1}^{n}+c_{2}\lambda_{2}^{n}$
with the boundary conditions $1=c_{1}+c_{2},\quad\abs a=c_{1}\lambda_{1}+c_{2}\lambda_{2}.$
It can be easily seen that $c_{1}>0$, $\lambda_{1}>\lambda_{2}>0$
since $\abs a^{2}>4\abs b^{2}$ by assumption of Lemma \ref{lem:detlemma}.
Now
\begin{align}
\underset{n\rightarrow\infty}{\text{lim}}\frac{1}{n}\lgbrac{\abs{A_{n}}} & =\underset{n\rightarrow\infty}{\text{lim}}\frac{1}{n}\lgbrac{c_{1}\lambda_{1}^{n}+c_{2}\lambda_{2}^{n}}\\
 & \overset{(a)}{=}\lgbrac{\lambda_{1}}\\
 & =\lgbrac{\abs a+\sqrt{\abs a^{2}-4\abs b^{2}}}-1.
\end{align}
The step $(a)$ follows because $\lambda_{1}>\lambda_{2}>0$ and $c_{1}>0$.

\section{Approximate capacity using n phase schemes \label{app:nphase_capacity}}

We have the following outer bounds from Theorem \ref{thm:apprx_capacity_fb}.
\begin{align}
R_{1},R_{2} & \leq\expect{\lgbrac{\abs{g_{d}}^{2}+\abs{g_{c}}^{2}+1}}\\
R_{1}+R_{2} & \leq\expect{\lgbrac{1+\frac{\abs{g_{d}}^{2}}{1+\abs{g_{c}}^{2}}}}+\expect{\lgbrac{\abs{g_{d}}^{2}+\abs{g_{c}}^{2}+2\abs{g_{d}}\abs{g_{c}}+1}}.
\end{align}
The above outer bound region is a polytope with the following two
non-trivial corner points:

\[
\left\{ \begin{array}{ccc}
R_{1} & = & \expect{\lgbrac{\abs{g_{d}}^{2}+\abs{g_{c}}^{2}+1}}\\
R_{2} & = & \expect{\lgbrac{1+\frac{\abs{g_{d}}^{2}}{1+\abs{g_{c}}^{2}}}}+\expect{\lgbrac{1+\frac{2\abs{g_{d}}\abs{g_{c}}}{1+\abs{g_{d}}^{2}+\abs{g_{c}}^{2}}}}
\end{array}\right\}
\]

\[
\left\{ \begin{array}{ccc}
R_{1} & = & \expect{\lgbrac{1+\frac{\abs{g_{d}}^{2}}{1+\abs{g_{c}}^{2}}}}+\expect{\lgbrac{1+\frac{2\abs{g_{d}}\abs{g_{c}}}{1+\abs{g_{d}}^{2}+\abs{g_{c}}^{2}}}}\\
R_{2} & = & \expect{\lgbrac{\abs{g_{d}}^{2}+\abs{g_{c}}^{2}+1}}
\end{array}\right\} .
\]
We can achieve these rate points within $2+3c_{JG}$ bits per channel
use for each user using the $n$-phase schemes since
\begin{align}
\left(R_{1},R_{2}\right) & =\left(\lgbrac{1+SNR+INR}-2-3c_{JG},\expect{\log^{+}\left[\frac{\abs{g_{d}}^{2}}{1+INR}\right]}\right)\\
\left(R_{1},R_{2}\right) & =\left(\expect{\log^{+}\left[\frac{\abs{g_{d}}^{2}}{1+INR}\right]},\lgbrac{1+SNR+INR}-2-3c_{JG}\right).
\end{align}
 are achievable and since using Jensen's inequality
\begin{equation}
\expect{\lgbrac{\abs{g_{d}}^{2}+\abs{g_{c}}^{2}+1}}\leq\lgbrac{1+SNR+INR}.
\end{equation}
The only important point left to verify is in the following claim.
\begin{claim}
$\expect{\lgbrac{1+\frac{\abs{g_{d}}^{2}}{1+\abs{g_{c}}^{2}}}}+\expect{\lgbrac{1+\frac{2\abs{g_{d}}\abs{g_{c}}}{1+\abs{g_{d}}^{2}+\abs{g_{c}}^{2}}}}-\expect{\log^{+}\left[\frac{\abs{g_{d}}^{2}}{1+INR}\right]}\leq2+c_{JG}$\label{claim:am_gm}
\end{claim}
\begin{IEEEproof}
We have $\frac{2\abs{g_{d}}\abs{g_{c}}}{\abs{g_{d}}^{2}+\abs{g_{c}}^{2}}\leq1$
due to AM-GM inequality. Hence,
\begin{equation}
\expect{\lgbrac{1+\frac{2\abs{g_{d}}\abs{g_{c}}}{1+\abs{g_{d}}^{2}+\abs{g_{c}}^{2}}}}\leq1.
\end{equation}
Also
\begin{equation}
\expect{\lgbrac{1+\frac{\abs{g_{d}}^{2}}{1+\abs{g_{c}}^{2}}}}\leq\expect{\lgbrac{1+\frac{\abs{g_{d}}^{2}}{1+INR}}}+c_{JG}
\end{equation}
using logarithmic Jensen's gap property. Hence, it only remains to
show $\lgbrac{1+\frac{\abs{g_{d}}^{2}}{1+INR}}-\log^{+}\left[\frac{\abs{g_{d}}^{2}}{1+INR}\right]\leq1$
to complete the proof.

If $\log^{+}\left[\frac{\abs{g_{d}}^{2}}{1+INR}\right]=0$ then $\frac{\abs{g_{d}}^{2}}{1+INR}\leq1$
and Hence, $\lgbrac{1+\frac{\abs{g_{d}}^{2}}{1+INR}}\leq\lgbrac 2=1$.

If $\log^{+}\left[\frac{\abs{g_{d}}^{2}}{1+INR}\right]>0$ then $\frac{\abs{g_{d}}^{2}}{1+INR}>1$
and Hence, again
\begin{equation}
\lgbrac{1+\frac{\abs{g_{d}}^{2}}{1+INR}}-\log^{+}\left[\frac{\abs{g_{d}}^{2}}{1+INR}\right]=\lgbrac{1+\frac{1+INR}{\abs{g_{d}}^{2}}}<1.
\end{equation}
\end{IEEEproof}

\section{Analysis for the 2-tap fading ISI channel\label{app:fading_isi_channel}}

We have for the outer bound
\begin{align}
n\brac{R-\epsilon_{n}} & \leq I\brac{Y^{n},g_{d}^{n},g_{c}^{n};W}\\
 & =I\brac{\rline{Y^{n};W}g_{d}^{n},g_{c}^{n}}\\
 & =h\brac{\rline{Y^{n}}g_{d}^{n},g_{c}^{n}}-h\brac{Z^{n}}\\
 & \leq\sum h\brac{\rline{Y_{i}}g_{d,i},g_{c,i}}-h\brac{Z^{n}}\\
 & \overset{\brac a}{\leq}\sum\expect{\lgbrac{1+P_{i}\abs{g_{d}}^{2}+P_{i-1}\abs{g_{c}}^{2}+2\abs{g_{d}}\abs{g_{c}}\sqrt{P_{i}P_{i-1}}}}\\
 & \leq\sum\brac{\expect{\lgbrac{1+P_{i}\abs{g_{d}}^{2}+P_{i-1}\abs{g_{c}}^{2}}}+1},
\end{align}
where $\brac a$ is using $P_{i}$ as the power for $i^{\text{th}}$
symbol and using Cauchy Schwarz inequality to bound $\abs{\expect{X_{i}X_{i-1}}}\leq\sqrt{P_{i}P_{i-1}}$.
Now using Jensen's inequality it follows that
\begin{align}
R-\epsilon_{n} & \overset{}{\leq}\expect{\lgbrac{1+\abs{g_{d}}^{2}+\abs{g_{c}}^{2}}}+1\\
 & \leq\lgbrac{1+SNR+INR}+1.
\end{align}
For the inner bound similar to the scheme in subsection \eqref{sec:FF_IC_p2p_codes},
using Gaussian codebooks and $n$ phases we obtain that

\begin{equation}
R=\underset{n\rightarrow\infty}{\text{lim}}\frac{1}{n}\expect{\lgbrac{\frac{\abs{K_{{\bf Y}}(n)}}{\abs{K_{{\bf Y|X}}(n)}}}}
\end{equation}
is achievable, where $\boldsymbol{X}(n)$ is $n$-length Gaussian
vector with i.i.d $\mathcal{CN}\brac{0,1}$ elements and $\boldsymbol{Y}\brac n$
is generated from $\boldsymbol{X}(n)$ by the ISI channel (from Equation
\eqref{eq:isi_channel}). Here $\abs{K_{{\bf Y|X}}(n)}=\abs{K_{{\bf Z}}(n)}=1$
because $Z$ is AWGN. Hence,
\begin{equation}
R=\underset{n\rightarrow\infty}{\text{lim}}\frac{1}{n}\expect{\lgbrac{\abs{K_{{\bf Y}}(n)}}}
\end{equation}
is achievable. Hence, it follows that
\begin{equation}
R\geq\lgbrac{1+SNR+INR}-1-3c_{JG}
\end{equation}
is achievable due to Lemma \ref{lem:fading_matrix} and Lemma \ref{lem:detlemma}.
\end{document}